\documentclass{aa}
\usepackage{natbib}
\bibpunct{(}{)}{;}{a}{}{,} % to follow the A&A style 
\usepackage{graphicx}
\usepackage{txfonts}
\usepackage{color}
\newcommand{\beq}{\begin{equation}}
\newcommand{\eeq}{\end{equation}}
\newcommand{\beqa}{\begin{eqnarray}}
\newcommand{\eeqa}{\end{eqnarray}}

\begin{document}
   \title{Flow instabilities of magnetic flux tubes}

   \subtitle{IV. Flux storage in the solar overshoot region}

   \author{E. I\c{s}\i k$^{1,3}$ \and V. Holzwarth$^{2,3}$}

   \institute{
	Department of Mathematics and Computer
	Science, \.Istanbul K\"ult\"ur University, Atak\"oy Campus,
	Bak\i rk\"oy 34156, \.Istanbul, Turkey \and 
	Fraunhofer-Institut f\"ur Kurzzeitdynamik, Ernst-Mach-Institut, 
	Eckerstra\ss e 4, 79104 Freiburg im Breisgau, Germany
	\and 
        Max-Planck-Institut f\"ur Sonnensystemforschung,
        Max-Planck-Stra\ss e 2, 37191 Katlenburg-Lindau, Germany \\	
             \email{e.isik@iku.edu.tr, volkmar.holzwarth@emi.fraunhofer.de}
             }
   \date{Received 3 July 2009 / Accepted 22 September 2009}

\abstract
{Flow-induced instabilities of magnetic flux tubes are relevant 
to the storage of magnetic flux in the interiors of stars with 
outer convection zones. The stability of magnetic fields in 
stellar interiors is of importance to the generation and 
transport of solar and stellar magnetic fields. }
{We consider the effects of material flows on the dynamics of 
toroidal magnetic flux tubes located close to the base of the 
solar convection zone, initially within the overshoot region. 
The problem is to find the physical conditions in which magnetic 
flux can be stored for periods 
comparable to the dynamo amplification time, which is of the order 
of a few years.}
{We carry out nonlinear numerical simulations to investigate the 
stability and dynamics of thin flux tubes subject to 
perpendicular and longitudinal flows. 
We compare the simulations with the results of simplified analytical 
approximations.}
{The longitudinal flow instability induced by the aerodynamic drag force 
is nonlinear in the sense that the growth rate depends on the perturbation 
amplitude. This result is consistent with the predictions 
of linear theory. 
Numerical simulations without friction show that 
nonlinear Parker instability can be triggered below the linear 
threshold of the field strength, when the difference in 
superadiabaticity along the tube is sufficiently large. 
A localised downflow acting on a toroidal tube in the overshoot 
region leads to instability depending on the parameters 
describing the flow, as well as the magnetic field strength. 
We determined ranges of the flow parameters for which a 
linearly Parker-stable magnetic flux tube is stored in the 
middle of the overshoot region for a period comparable to the dynamo 
amplification time. }
{The longitudinal flow instability driven by frictional interaction of 
a flux tube with its surroundings is relevant to determining the storage 
time of magnetic flux in the solar overshoot region. 
The residence time for magnetic flux tubes with $2\times 10^{21}$Mx 
in the convective overshoot layer is comparable to the dynamo 
amplification time, provided that the average 
speed and the duration of the downflow do not exceed about 
50~m~s~$^{-1}$ 
and 100~days, respectively, and that the lateral extension of 
the flow is smaller than about $10^\circ$. 
}


\titlerunning{Flow instabilities of magnetic flux tubes IV}
\authorrunning{E. I\c{s}\i k \& V. Holzwarth}
\maketitle
%
%________________________________________________________________

\section{Introduction}
\label{sec:intro}
Observations of large solar active regions are indicative of an 
organised subsurface magnetic field along the azimuthal 
(east-west) direction, with 
opposite polarity orientations (from east to west or vice versa) 
in the northern and southern 
hemispheres. Emerging magnetic structures are in a filamentary state, 
in the form of magnetic flux concentrations (flux tubes) of various 
sizes (e.g., sunspots, pores). 
Flux tubes rise in the convection zone, emerge at the surface, and 
form bipolar magnetic regions, which follow 
polarity rules (Hale's law) and systematic tilt angles (Joy's law). 

Theoretical studies indicate that weak magnetic fields are 
transported by flux expulsion \citep{msch84} and convective 
pumping \citep{tobias01} to the lower 
boundary of the convection zone, 
where the toroidal magnetic field is amplified by radial 
and latitudinal velocity 
shear in the solar tachocline \citep[see e.g.,][]{sis06}. 
The stably stratified lower convection zone, in particular the 
convective overshoot layer, 
is a likely location for the generation and 
storage of the large-scale azimuthal magnetic flux. 
Numerical studies simulating a layer of horizontal magnetic 
field in the bottom of the solar convection zone indicate that 
an initially uniform field underlying a field-free layer 
leads to the formation of magnetic flux tubes by magnetic 
Rayleigh-Taylor instability \citep[e.g.,][]{fan01}. 
Following its formation, a toroidal magnetic flux tube can reach 
a mechanical equilibrium close to the bottom of the convection 
zone \citep{mi92}. 
The emergence of magnetic flux tubes driven by magnetic buoyancy 
and the properties of active regions (low-latitude emergence, 
tilt angles, proper motions of sunspots) require azimuthal flux 
densities of the order of $10^5$~G in the overshoot region 
\citep[][]{dsilva93,fan94,cale95,msch96}. 
The possibilities for the generation of these flux densities have been 
reviewed by \citet{schrempel02}, \citet{mschafm03}, and \citet{afm07}. 

The average duration of the solar activity cycle is about 11 years 
and the total magnetic flux emerging within one cycle 
ranges between orders of $10^{24}-10^{25}$~Mx. 
The amplification of the toroidal magnetic field 
to its maximum strength in about half an activity cycle raises the 
following question: how can the toroidal flux be stored stably 
for at least a few years in the dynamo amplification region? 
A related problem concerns the feedback of magnetic flux loss 
on the rate of toroidal flux generation. 
To more clearly understand magnetic flux generation and 
storage in the convective overshoot region, it is important to determine 
and constrain the effects of flows on the stability and 
dynamics of magnetic flux tubes. 
Here, we focus on: (a) the nonlinear development of 
flow-induced flux tube instabilities, which can be dynamically 
significant in the course of toroidal field amplification; and 
(b) effects of perpendicular flows on the storage of toroidal flux tubes. 
Consequently, some of the flow 
properties prevailing in the overshoot region can also be constrained, 
by requiring that flux tubes with field strengths of up to a few times 
$10^4$~G are stored in the overshoot region for about a few years. 

We consider the possibility that the 
friction-induced instability \citep[][]{pap2,pap3} 
leads to flux loss from the overshoot layer 
for field strengths below the Parker instability limit. 
We refer to such Parker-stable flux tubes as ``sub-critical'' 
throughout the paper. We investigate effects of flows on 
magnetic flux tubes in mechanical equilibrium to obtain 
quantitative estimates, and to answer the following 
question: 
how strongly do ($i$) finite-amplitude perturbations of flux tubes 
by perpendicular flows \citep[see also][paper~I]{pap1} and ($ii$) the 
frictional interaction of the tube 
with its surroundings limit the residence time of flux tubes in the 
overshoot region? 

We approach the problem by considering 
the nonlinearity of friction-induced instability 
(Sect.~\ref{sec:friction}),
including a discussion of the effects of finite perturbations 
on the magnetic buoyancy instability (Sect.~\ref{ssec:delta}), and 
determine the ranges of flow parameters that allow the storage of 
sub-critical flux tubes subject to radial flows (Sect.~\ref{sec:flows}). 

The analyses and simulations were carried out in the framework 
of the thin flux tube approximation \citep{spruit81}, which at present 
is the only 
existing approach that can deal with the small magnetically induced 
variations in density, pressure, and temperature corresponding to 
the high plasma $\beta$ ($>10^5$) of the deep solar convection zone. 

\section{Friction-induced instabilities}
\label{sec:friction}

The mechanical equilibrium of a toroidal flux tube in the solar convection 
zone requires that the plasma within the tube rotates faster than the 
surrounding medium. \citet{mi92} demonstrated that a toroidal flux tube 
rotating initially at the same rate as the external medium can 
reach this equilibrium by developing an internal prograde flow. 
The speed of this ``equilibrium'' flow depends on 
the magnetic field strength, depth, and latitude. 
A flux tube subject to non-axisymmetric perturbations (varying 
in azimuth) will develop MHD waves that propagate in prograde 
and retrograde directions. 
If the internal flow speed (in the rest-frame of the external medium) 
is greater than the phase speed of the slowest transversal retrograde 
wave mode, this wave mode will be advected in the prograde direction, 
while being amplified \citep[][hereafter paper~II]{pap2}. 
The critical flux density for the instability 
is lower than that of the Parker instability. 
\citet[][hereafter paper~III]{pap3} extended the analysis in paper~II 
to toroidal flux tubes, thus including the effect of magnetic curvature 
force and rotation. 

The linear analyses in papers II and III considered Stokes-type 
friction, which is linear in perpendicular velocity. 
To test the results of the linear stability analysis 
in the nonlinear context, we shall 
consider the drag force, which is quadratic in velocity. 
This test can be implemented by carrying out fully nonlinear numerical 
simulations of flux tubes subject to non-axisymmetric perturbations. 

\begin{figure}%[ht!]
%\centering
\resizebox{\hsize}{!}{\includegraphics[width=.9\linewidth]{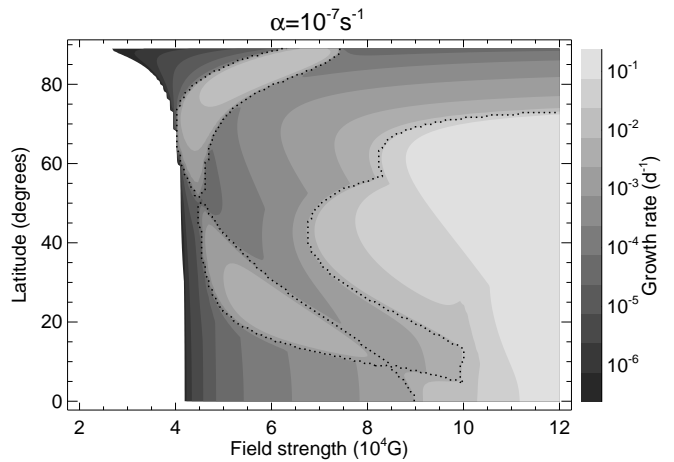}}
\caption{Stability diagram for toroidal flux tubes located in the middle of the 
solar overshoot region. The shaded areas indicate 
unstable configurations with the grey scale representing the growth rate. 
Dotted lines outline regions of Parker instability.}
\label{fig:stabmid}
\end{figure}

\subsection{Linear results}
\label{ssec:linear}

The hydrodynamic drag force exerted by a steady flow with velocity 
${\bf v}_\perp$ perpendicular to a straight cylinder is given by 
\begin{eqnarray}
	{\bf F}_{\rm D} = 
	-C_{\rm D}\frac{\rho_{e0}\varv_\perp{\bf v}_\perp}{\pi R_{\rm t}}, 
	\label{eq:drag}
\end{eqnarray}
where $C_D$ is the drag coefficient, $\rho_{e0}$ 
is the density of the external medium neighbouring the 
equilibrium flux tube, and $R_{\rm t}$ is the 
cross-sectional radius of the equilibrium tube. 

The Stokes-type friction force per unit volume, 
\begin{eqnarray}
{\bf F}_{\rm St} = -\alpha\rho_{e0} {\bf v}_\perp,
\label{eq:fst}
\end{eqnarray}
where the constant $\alpha$ is the friction coefficient, 
was considered in papers~II and III to include the 
friction-type deceleration in the linearised equation of motion. 
In paper~II it was shown that the critical field strength for 
friction-induced instability hardly depends on the value of $\alpha$. 
Comparing Eq.~(\ref{eq:fst}) with Eq.~(\ref{eq:drag}), 
one finds that the friction coefficient is of the order 
\begin{eqnarray}
\alpha \sim \frac{\varv_\perp}{\pi R_{\rm t}},  
\label{eq:cst}
\end{eqnarray}
where the drag coefficient, $C_{\rm D}$, is set to be unity, because of 
the cylindrical cross-section and high Reynolds number \citep{batchelor67}. 

We consider a toroidal flux tube located in the convective 
overshoot region, parallel to the equatorial plane. 
For the stratification of the ambient medium, we use a model convection zone 
developed by \citet{skastix91}, which uses 
a non-local treatment of convection as described by \citet{shavsal73}. 
In the model, the overshoot region extends about $10^4$~km below the base 
of the convection zone, which is defined as the depth at which 
the convective energy flux changes its sign, at about 
$r=512$~Mm. 
The thickness of the overshoot layer corresponds to about 
20\% of the local pressure scale height. 
Throughout the paper, the terms \emph{bottom, middle, and top (levels) of the 
overshoot region} are used to refer to the radial 
positions at, respectively, 2000, 5000, and 8000~km above the lower 
boundary of the overshoot region, which is at a radius of $r=502$~Mm 
($\sim 0.72R_\odot$). 

The linear stability analysis of paper~III has provided 
growth rates of the friction-induced instability for toroidal flux tubes. 
In Fig.~\ref{fig:stabmid}, growth rates are shown as a function of the 
field strength ($B_0$) and latitude ($\lambda_0$) 
located in the middle of the overshoot region, 
for $\alpha=10^{-7}$~s$^{-1}$ and rigid solar rotation\footnote{Throughout 
the paper, the index ``0'' refers to quantities 
pertaining to the mechanical equilibrium state.}. In paper~III, 
the value of $\alpha$ was chosen by determining the 
average perpendicular velocity of the mass elements of the flux tube 
during the initial stages of nonlinear numerical simulations 
for a flux tube with $\Phi\simeq 3\times 10^{21}$~Mx. 
The friction-induced instability sets in for $B_0\gtrsim 4\times 10^4$~G. 
The onset of Parker instability is in the interval 
$6\times 10^4 \lesssim B_0\lesssim 10^5$~G at low latitudes. 

\subsection{Nonlinear simulations}
\label{ssec:nonlinsims}

We carried out numerical simulations of a toroidal flux tube 
using a semi-implicit finite-difference scheme developed 
by \citet{moreno86}, that was
extended to three dimensions and spherical geometry by \citet{cale95}. 
The numerical procedure is based on 
the equations of ideal magnetohydrodynamics in the framework of 
the thin flux tube approximation \citep{spruit81}, in the form 
given by \citet[][]{afmsch93,afmsch95}. 
In the numerical scheme, 
the flux tube is described by a string of Lagrangian mass elements, 
which move in three dimensions under the effects of various body 
forces. 
The equation of motion for the material inside the toroidal flux tube, 
in a reference frame rotating with the angular velocity of the tube, 
${\bf \Omega}$, is written as 
\begin{eqnarray}
    \rho_i\frac{D{\bf v}_i}{Dt} &=&
    -\nabla\Big(p_i+\frac{B^2}{8\pi}\Big) + 
    \frac{({\bf B\cdot\nabla}){\bf B}}{4\pi} \nonumber \\
    & & + \rho_i\big[{\bf g - \Omega\times(\Omega\times r)}\big] + 
    2\rho_i{\bf v}_i{\bf\times\Omega} + {\bf F}_{\rm D},
\end{eqnarray}
where $D/Dt\equiv\partial/\partial t+{\bf v}\cdot\nabla$ 
is the Lagrangian derivative, the subscript $i$ denotes 
quantities inside the flux tube (the subscript $e$ denotes external 
quantities in the following). The terms on 
the right hand side of the equation are, respectively, the total 
pressure force (gas and magnetic), 
magnetic tension force, effective gravity (including the 
centrifugal force), Coriolis force, and the hydrodynamic drag force, 
which is given by Eq.~(\ref{eq:drag}). 
We assume rigid rotation in the solar interior.
Although rotational velocity shear 
can be incorporated into the framework of our model 
\citep{afmsch95,cale95}, we disregard it
for the sake of clarity, to focus on the fundamental effects 
of radial flows on flux tube dynamics. 
\citet{cale95} found that
external shear does not have a significant effect on the stability
properties and dynamics of flux tubes in the Sun.

\begin{figure}%[ht!]
%\centering
\resizebox{\hsize}{!}
{\includegraphics[width=.9\linewidth]{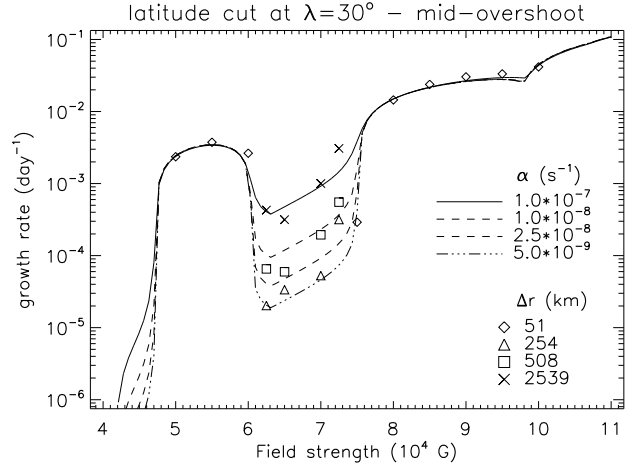}}
	\caption{Comparison of the analytical solutions for the growth rate of 
	the friction-induced instability (lines) with the numerical simulations 
	(symbols) for flux tubes at the middle of the overshoot region and 
	$30^\circ$ latitude. The initial perturbation amplitude ($\xi$), 
	and the Stokes drag coefficient ($\alpha$) are indicated in 
	the legend. }
	\label{fig:lamslice}
\end{figure}

The initial value for the 
cross-sectional radius of each flux tube is taken to be $R_{\rm t}=1000$~km 
for all simulations. 
For $B_0=10^5$~G, this corresponds to a magnetic flux of about 
$3\times 10^{21}$~Mx, which is typical of a bipolar magnetic region 
of moderate size on the solar surface. 

The flux tube in mechanical equilibrium is initially perturbed by 
small displacements in three dimensions. 
The azimuthal dependencies of the initial perturbations in 
each coordinate $k$ were taken to be in the form
\begin{eqnarray}
\xiup_k &=& \sum_{m=1}^5 \xi_{k0}\sin m\phi_0,
\label{eq:sinepert}
\end{eqnarray}
where the amplitudes of all modes are equal. 
In cases of instability, the growth rate of the 
perturbation was determined by an exponential fit to the radial 
location of the top of the growing loop. 
Each fit was limited to the time interval corresponding to the 
initial exponential growth phase of the instability. 

We compare the linear results for 
a range of friction parameters $\alpha$, with the results of nonlinear 
simulations in Fig.~\ref{fig:lamslice}, which shows the growth rates 
of the friction-induced 
instability as a function of the magnetic field strength, for $30^\circ$ 
latitude in the middle of the overshoot region, for chosen values of 
$\alpha$. 
The two plateaus where the curves for the linear solutions 
converge correspond to Parker-unstable regions according to the 
linear analysis (regions enclosed by 
dots in Fig.~\ref{fig:stabmid}). Thus the curve for 
$\alpha=10^{-7}~{\rm s}^{-1}$ in Fig.~\ref{fig:lamslice} 
can be seen as a horizontal cut through Fig.~\ref{fig:stabmid} at 
$\lambda_0=30^\circ$. 
Each set of numerical simulations were performed with a fixed 
value of the total initial perturbation amplitude in the radial 
direction, $\Delta r \equiv 2|\xiup|$, ranging between 51~km and 2539~km.
The values of $\alpha$ were chosen such that 
the curves correspond to the range of growth rates found in the simulations. 
The simulations exhibit an overall 
similarity with the linear results in terms of the field 
strength dependence of the growth rate. The linear growth rate of the 
instability based on a fixed value of $\alpha$ corresponds approximately to 
a certain perturbation amplitude, $\Delta r$, as a function of $B_0$ in 
the linearly Parker-stable and frictionally unstable intermediate regime. 

The coefficient $\alpha$ is a measure of the strength of frictional 
coupling of the oscillating flux tube with the surrounding medium. 
For $\alpha\lesssim 10^{-5}~{\rm s}^{-1}$, 
the growth rate is proportional to $\alpha$, 
because the frictional coupling facilitates the amplification of 
perturbations. For $\alpha\gtrsim 10^{-5}~{\rm s}^{-1}$, on the 
other hand, the growth rates begin to decrease with increasing 
$\alpha$, because too large friction impedes perpendicular movements 
of the flux tube and thus the development of overstability. 
The $\alpha$ values chosen in Fig.~\ref{fig:lamslice} correspond to 
the regime in which friction has a destabilising effect. 
The reason for the correspondance between $\alpha$ 
and $|\xiup|$ is the quadratic dependence of the drag 
force on the perpendicular velocity: a larger initial 
perturbation leads to a larger perpendicular velocity 
and thus to a larger $\alpha$, meaning a stronger Stokes-frictional 
coupling. The nonlinear growth of the instability 
is shown in a supplemented animation (see the online appendix, 
Fig.~\ref{fig:frict_inst}), for a flux tube 
with $B_0=7\times 10^4$~G, $\lambda_0=30^\circ$, and $\Delta r=5508$~km. 

\begin{figure}%[ht!]
%\centering
\resizebox{\hsize}{!}
{	\includegraphics[width=.9\linewidth]{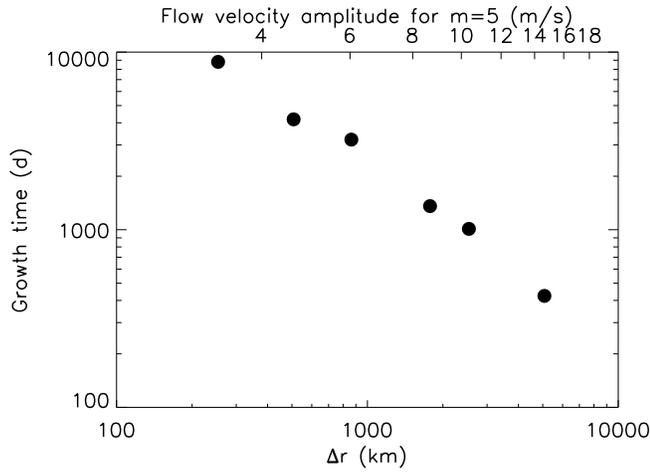} }
	\caption{Growth time as a function of the displacement amplitude, 
	$B_0=7\times 10^4$~G, $\lambda_0=30^\circ$, at the middle of the overshoot 
	region. The upper horizontal axis shows the flow velocity amplitude 
	that leads to the displacement $\Delta r$ according to the analytical 
	approximation discussed in Sect.~\ref{ssec:perlin}, for $m=5$ and 
	$|\varv_\perp|_{\rm max}=26$~m~s$^{-1}$. }
	\label{fig:gr-amp1}
\end{figure}

The variation in growth time as a function of perturbation amplitude 
for $B_0=7\times 10^4$~G (see Fig.~\ref{fig:lamslice}) is shown in 
Fig.~\ref{fig:gr-amp1}. It shows that the friction-induced instability 
grows faster for larger perturbations, following a power law. 

There are two reasons for the nonlinear behaviour of the instability, 
i.e., that the growth rate is proportional to the initial perturbation 
amplitude: (\emph{i}) $\alpha$ is proportional to $\varv_\perp$, i.e, 
the drag force 
\emph{is} nonlinear, (\emph{ii}) the spatial variation in the 
superadiabaticity, $\delta\equiv\nabla - \nabla_{\rm ad}$, 
of the external medium along the flux tube 
becomes increasingly effective with increasing $|\xiup|$, such that the 
excessive magnetic buoyancy at the highest location of the tube provides 
an additional upward acceleration. However, the latter ``$\delta$-effect'' 
is not likely 
to be the dominant source of nonlinearity, because otherwise 
there would be far poorer agreement between the linear and nonlinear 
results in Fig.~\ref{fig:lamslice}. 
This point will be investigated further in Sect.~\ref{ssec:delta}. 

We carried out additional simulations for linearly Parker-stable 
configurations with a perturbation of $m=1$, by 
varying the amplitude. The results are given in Table~\ref{tab:sims}. 
In all cases but one for $B_0=8\times 10^4$~G and $\lambda_0=60^\circ$, 
friction was taken into account. 
For $B_0=3.5\times 10^4$~G, the flux tube is linearly stable 
(see Fig.~\ref{fig:stabmid}), and is also stable to the finite perturbations 
applied in the simulations. For stronger fields the friction-induced 
instability sets in, the 
growth rate then being proportional to the perturbation amplitude. 
For a given latitude and perturbation amplitude, a higher $B_0$ leads 
to a more rapid growth, as predicted by the linear approach. 
Owing to nonlinear effects, the initial $m=1$ perturbation 
initiates perturbations of higher-order modes, in particular if 
the initial amplitude is large. In the case of unstable flux tubes, 
the eigenmode with the shortest growth time dominates the evolution 
and causes an increase in the top position of the flux tube. 
During its early evolution, this rise is approximately exponential 
and can be characterised by the e-folding times given in Table~\ref{tab:sims}.
In those cases marked with asterisks, 
the top position of the perturbed flux tube decreases with time. 
Since each eigenmode decays on its own 
individual damping time and modes may interact nonlinearly, a unique 
e-folding timescale of the overall decay of the top position is not 
possible. After all, in these cases we find that the top position 
decreases to about 
one tenth of its original value within $10^3 - 10^4$~days. 
The configuration $B_0=8\times 10^4$~G and $\lambda_0=60^\circ$ is 
very close to the Parker instability boundary (see Fig.~\ref{fig:stabmid}), 
and the result for a sufficiently large perturbation 
is a rapid growth of an instability induced by 
the steep $\delta$-gradient of the external medium along the tube, which we 
consider in the next section. 

\begin{table}%[ht]
\centering
\caption{Growth times for various finite perturbations.}
\begin{tabular}{c c c c c}
\hline
$B$~($10^4$~G) & $\lambda_0$ ($^\circ$) & $\Delta r_0$ (km) & Friction & Growth time (d) \\ [0.5ex]
\hline\hline
3.5	&	10	&	1280	&	1	&	* \\
3.5	&	10	&	12800	&	1	&	* \\
\hline
3.5	&	60	&	1280	&	1	&	* \\
3.5	&	60	&	15090	&	1	&	* \\
\hline
5.0	&	10	&	1509	&	1	&	* \\
\hline
5.0	&	60	&	1509	&	1	&	99000 \\
\hline
7.0	&	10	&	13	&	1	&	* \\
7.0	&	10	&	128	&	1	&	79500 \\
7.0	&	10	&	1280	&	1	&	6950 \\
7.0	&	10	&	6400	&	1	&	789 \\
\hline
7.0	&	30	&	15	&	1	&	* \\
7.0	&	30	&	151	&	1	&	* \\
7.0	&	30	&	1510	&	1	&	7790 \\
7.0	&	30	&	7550	&	1	&	512 \\
7.0	&	30	&	15090	&	1	&	112 \\
\hline 
7.0	&	60	&	15	&	1	&	* \\
7.0	&	60	&	151	&	1	&	13000 \\
7.0	&	60	&	1510	&	1	&	1020 \\
7.0	&	60	&	7550	&	1	&	126 \\
\hline
8.0	&	60	&	15	&	1	&	9370 \\
8.0	&	60	&	151	&	1	&	872 \\
8.0	&	60	&	1510	&	1	&	98 \\
\hline
8.0	&	60	&	3020	&	0	&	* \\
8.0	&	60	&	7550	&	0	&	42 \\ 
\hline
\end{tabular}
\label{tab:sims}
\end{table}

\subsection{Effect of stratification}
\label{ssec:delta}

\begin{figure}
\centering
\includegraphics[width=\linewidth]{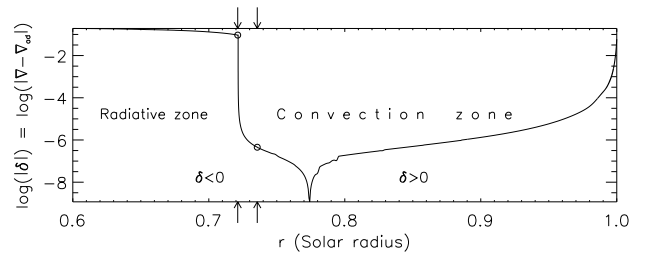}
\caption{Radial profile of the superadiabaticity in the outer 
40 per cent of the Sun, according to the solar model used 
\citep{skastix91}. 
The arrows and the corresponding circles mark the boundaries of the 
overshoot region.}
\label{fig:delta}
\end{figure}

The most important external 
quantity in determining the stability of a toroidal flux tube 
in the overshoot region is the superadiabaticity, $\delta$. 
Figure~\ref{fig:delta} shows the 
radial profile of $\log|\delta|$ in the outer 40\% of the solar 
interior, according 
to the model under consideration \citep{skastix91}. 
From the upper radiative 
zone to the lower convection zone, the superadiabaticity 
increases from negative values and changes its sign at about 
$r=0.77R_{\sun}$. It increases by a factor of about 
$2\times 10^5$ through the overshoot layer. 
We assume that a toroidal flux tube is deformed in such a way that 
it extends between the boundaries of the overshoot region. 
The parts of the tube that extend to higher layers (smaller $|\delta|$) 
will experience larger buoyancy force and thus be destabilised, whereas 
the parts that extend to deeper layers will be stabilised owing to 
sufficiently negative superadiabaticity. 
These nonlinear effects may lead to the formation of a rising flux loop.

\begin{figure}%[ht!]
%\centering
\resizebox{\hsize}{!}
{	\includegraphics[width=.9\linewidth]{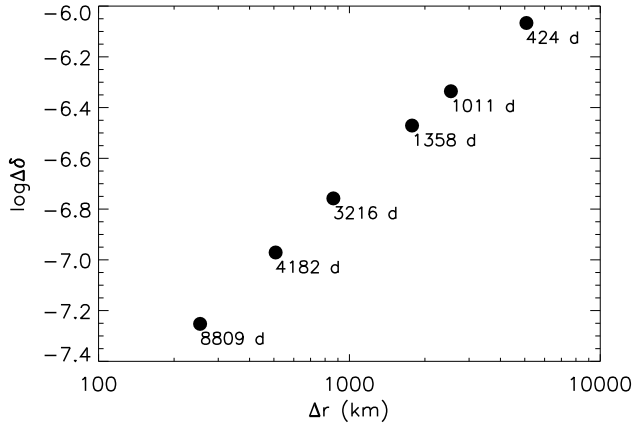} }
	\caption{The difference in superadiabaticity between the
        surroundings
	of the top and bottom parts of perturbed flux tubes, which are
        initially located at the middle overshoot zone with
        $\lambda_0=30^\circ$ and $7\times 10^4$~G, the same initial
        conditions as in Fig.~\ref{fig:gr-amp1}. The drag force has 
        been taken into account. The numbers indicate growth times of 
        the instability for each simulation.}
	\label{fig:subad_pert}
\end{figure}

Figure~\ref{fig:subad_pert} shows 
the initial superadiabaticity difference, $\Delta\delta$, 
between the surroundings of the top and bottom parts of the flux tube 
as a function of the initial perturbation amplitude, $\Delta r$,
for the case presented in Fig.~\ref{fig:gr-amp1}
(Sect.~\ref{ssec:nonlinsims}).
We note that the relative difference $\Delta\delta/\delta$ is lower 
than unity ($\log\delta\approx -6.0$).

To check whether a variation in $\delta$ is responsible for 
the nonlinear dependence of the growth rate on the perturbation amplitude, 
we carried out simulations with the same initial conditions as 
given above (Fig.~\ref{fig:subad_pert}), but without the drag force, 
to avoid a 
mixing with the frictional instability, thus isolating the $\delta$-effect. 
No instability was found within the considered 
range of perturbation amplitudes. 
This indicates that the $\delta$-effect is not responsible for the 
nonlinearity of the frictional instability for $\log\Delta\delta<-6.0$. 

\begin{figure}%[ht!]
%\centering
\resizebox{\hsize}{!}
{	\includegraphics[width=.9\linewidth]{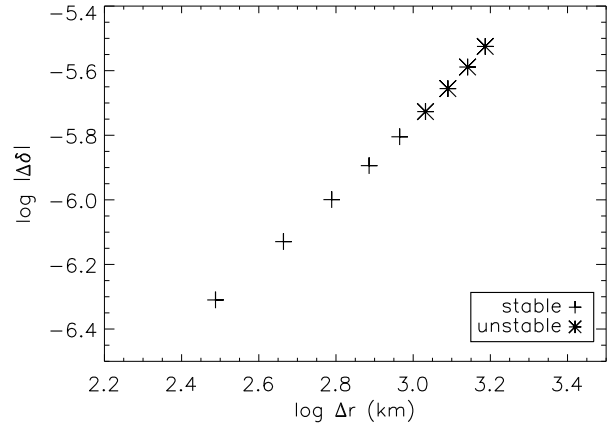} }
	\caption{Same as Fig.~\ref{fig:subad_pert}, 
	for $B_0=12.6\times10^4$ G, $\lambda_0=60^\circ$, at the bottom of the 
	overshoot region, without the drag force. When the initial 
	perturbation amplitude exceeds $\sim 10^3$~km, the flux tube 
	becomes unstable owing to buoyancy differences along the tube. 
	}
	\label{fig:stabtest_bot}
\end{figure}

In order to constrain the difference of superadiabaticity 
required for the $\delta$-effect to have a significant role in 
destabilising a flux tube, we carried out numerical simulations 
at the bottom of the overshoot layer, where the radial gradient 
of $\delta$ is steeper than in the upper overshoot 
region. Two sets of simulations were carried out for 
latitudes $\lambda_0=60^\circ$ and $\lambda_0=20^\circ$ and 
$B_0=12.6\times10^4$~G, without the drag force. 
The two cases correspond to linearly Parker-stable configurations, 
very close to the instability boundary, which is about 2 and 4~kG 
larger for the high- and low-latitude cases, respectively. 
For both cases, we find instability 
for $\Delta r\gtrsim 1000$~km. 
Figure~\ref{fig:stabtest_bot} 
shows $\Delta\delta$ as a function of $\Delta r$ 
for the high-latitude case. 
A rising loop is formed for $\Delta r\gtrsim 1000$~km, 
which corresponds to $\log |\Delta\delta|\gtrsim -5.8$. 
The growth rates are about 60 days for 
$\Delta r\simeq 1076$~km and 20 days 
for $\Delta r\simeq 1537$~km, which are comparable with the rise time in the 
convection zone proper. The rising loop is triggered by
nonlinear effects owing to a sufficiently large magnetic buoyancy 
difference along the flux tube, rather than by overstability of interacting 
wave modes. 

To summarise the results obtained in this section, we find that a 
superadiabaticity 
difference of the order of $10^{-5}$ is required for a subcritical 
toroidal flux tube to become unstable in the overshoot layer, due to 
the $\delta$-effect. 

\subsection{The slingshot case}
\label{ssec:slingshot}

One may expect that rapidly rising flux loops can be 
formed by localised upflows at the top of the overshoot region. 
An interesting 
question is whether such an eruption of a small part of the flux tube 
can originate close to the top of the overshoot region. 
To test this possibility, we carried out a 
numerical simulation for a flux tube at the top of the overshoot region 
with $B_0=6\times 10^4$~G and $\lambda_0=40^\circ$, which is a 
linearly Parker-unstable configuration. The perturbation was 
applied with $|\xi_r|=1500$~km and $m=15$, which corresponds to 
an azimuthal extension of $12^\circ$ for each lifted portion. 
In the course of the simulation, the wave energy in high azimuthal 
modes is gradually 
transferred into lower-$m$ modes. The tube enters into the convection 
zone with $m=2$ mode, forming two large-scale loops. This is 
consistent with the prediction of the linear stability analysis. 

Small-scale loops can also originate in the bottom of the overshoot 
region, where the $\delta$-effect (Sect.~\ref{ssec:delta}) is 
significant. A slingshot effect can occur at the radiative zone 
boundary, if a sufficiently strong localised downflow pushes a small 
part of the tube downward. Subsequently, the submerged part would 
be ejected upwards owing to strong buoyancy. 
We defined initial conditions describing localised 
downward perturbations of Gaussian shape with the azimuthal extension 
ranging from about $9^\circ$ to $18^\circ$ ($10\leq m \leq 20$) 
at the bottom of the overshoot region. At the beginning of the simulation, 
the submerged part of the tube rises rapidly from the radiative zone 
boundary. 
However, because of the drag force, which is proportional to 
the square of the perpendicular velocity, it has already been rapidly 
decelerated within the 
overshoot region and its translational kinetic energy has been partly 
transferred into MHD waves propagating along the tube. 

\section{Displacement of a toroidal flux tube by radial flows}
\label{sec:flows}

We have so far considered the dynamics of toroidal 
flux tubes 
based on the assumption of spatially perturbed initial configurations, 
without an explicit external driving force.
In the following sections, we investigate the effects 
of external flows perpendicular to the tube axis, which displace 
the tube owing to the drag force, given by Eq.~(\ref{eq:drag}). 
The purposes are (1) to quantify the effects of external flows on 
the subsequent displacement, and 
(2) to test the possibility of storing a toroidal magnetic flux 
tube with $\Phi\sim 10^{21}-10^{22}$~Mx in the solar overshoot region 
for a few years, which is comparable to the dynamo amplification 
time. 
We examine the effects of spatially periodic radial flows 
in Sect.~\ref{ssec:perlin}, 
and of localised downflows in Sect.~\ref{ssec:localised}. 

\subsection{Azimuthally periodic flow: linear analysis}
\label{ssec:perlin}

We assume that a toroidal flux tube is deformed by azimuthally 
periodic radial flows to such an extent that the drag force is balanced 
by buoyancy and magnetic tension. 
By solving the linearised equations of motion 
for the perturbations of a thin flux tube, 
we can derive an analytical expression relating 
flux tube parameters to flow parameters. 
This relation allows us to estimate the conditions 
in which the deformation becomes 
so large that parts of the flux tube enter radiative and/or 
convection zones. These deformations can destabilise a flux tube 
in the overshoot region, e.g., by the $\delta$-effect. 

\subsubsection{Solution procedure}
\label{ssec:sol}

\begin{figure}
	\includegraphics[width=.9\linewidth]{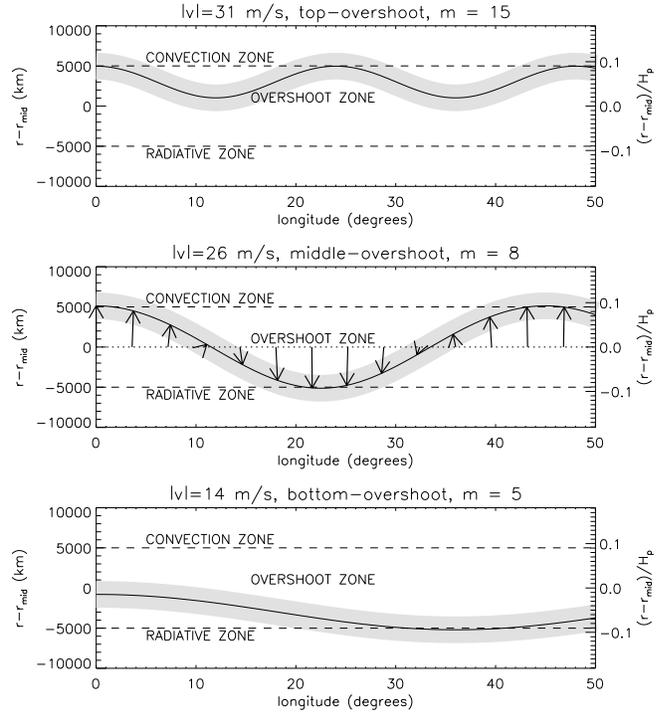}
        \caption{Geometry of a flux tube deformed by azimuthally 
	periodic flows, at three initial locations (solid lines): 
	(from top to bottom of the figure) the top, the middle, and the 
	bottom of the overshoot region. The latitude is 
	$30^\circ$ and the field strength is $6\times 10^4$~G. The distance 
	from the middle of the overshoot region 
	in kilometres is shown on the left axis, and in units of local 
	pressure scale height (at $r=r_{\rm mid}$) on the right axis. 
	The azimuthal wavenumber is chosen such that the troughs or crests 
	of the tube partially enter the convection or the radiative zones. 
	The radius of the tube is 1000~km. Arrows denote the relative 
	strength of the resultant perturbation, which is led by the external 
	flow. }
        \label{fig:stabg}
\end{figure}

We adopt the linearised equations of thin magnetic flux
tubes as given by 
\citet{afmsch93,afmsch95} and apply the drag force exerted by an 
azimuthally periodic flow of the form 
\begin{eqnarray}
F_{\rm D}(\phi)  &=& 
	\frac{\rho_{e0}\varv_{\perp 0}^2}{\pi R_{\rm t}}\exp(im\phi). 
\label{eq:vel}
\end{eqnarray}
The azimuthal wavenumber, $m$, measures the azimuthal extension 
of the perpendicular flow, $\pi/m$. 
We assume that the flux tube reaches a stationary equilibrium state, 
where the drag force is balanced by buoyancy and magnetic tension forces, 
so that the time derivative in the momentum equation vanishes. 
The components of the resulting linearised equations of motion for 
perturbations in steady state are given in 
Appendix~\ref{sec:ftradial}, 
Eqs.~(\ref{eq:momeqout1})-(\ref{eq:momeqout3}). 
For an azimuthally periodic flow, 
the components of the displacement as a function of azimuth 
in cylindrical coordinates are 
\begin{eqnarray}
\xi_R = \Re{\rm e}\hat{\xi}_R \cdot \cos m\phi, \nonumber \\
\xi_\phi = -\Im{\rm m}\hat{\xi}_\phi \cdot \sin m\phi, \nonumber \\
\xi_z = \Re{\rm e}\hat{\xi}_z \cdot \cos m\phi,
\label{eq:sool}
\end{eqnarray}
where the complex amplitudes $\hat{\xi}_R,\hat{\xi}_\phi,\hat{\xi}_z$ 
are given by Eqs.~(\ref{eq:sool1})-(\ref{eq:sool3}). 
The phase difference between the azimuthal displacement 
$\xi_\phi$ and the spherical radial displacement $\xi_r$ 
is $\pi/2$. The azimuthal perturbation leads to diverging flows around 
tube crests and converging flows around the troughs of the tube, 
to restore hydrostatic equilibrium (see Fig.~\ref{fig:stabg}, 
middle panel). 
Such a flow pattern increases the density deficit in the tube crests, 
thus it has a destabilising effect. 
The azimuthal perturbation does not have a significant effect on the 
radial position of the tube crests, owing to (\emph{i}) the relatively 
small ratio $\xi_\phi/\xi_R$ and (\emph{ii}) the phase relation between 
$\xi_R$ and $\xi_\phi$. 

In the present analysis, we assume that any displacement 
of mass elements along the tube is negligible 
\emph{before} the stationary equilibrium is reached. 
This process 
involves nonlinear variations in density, internal flow speed, 
and magnetic field strength as functions of azimuth and time, 
which are outside the scope of our linear approach. 
We present the justification of this assumption and discuss its 
limitations in Appendix~\ref{sec:assumption}. 

We begin the analysis by finding the extreme values 
of the velocity amplitude, that displaces parts of the flux tube to 
the boundaries of the overshoot layer. 
Figure~\ref{fig:stabg} shows the spherical perturbation $\xi_r$ as a 
function of $\phi$ at three depths in the overshoot region for 
a flux tube with $B_0=6\times 10^4$~G, $\lambda_0=30^\circ$, and 
$R_t=1000$~km. 
The velocity amplitudes corresponding to each of these three depths 
were adopted from the convection zone model of \citet{skastix91}. 
The azimuthal 
wavenumber were chosen such that in each case the crests or the 
troughs of the perturbed tube come very close to the boundaries of 
the overshoot layer. We call this azimuthal wavenumber $m_{\rm crit}$ 
and the resulting perturbation the {\it critical perturbation}. 
At the bottom of the overshoot region, the flow must be very extended 
($m\lesssim 5$) to destabilise the flux tube. 
In higher layers of the overshoot region, narrower flows ($m\lesssim 8$ 
and $m\lesssim 15$) can have a destabilising effect. 
The critical azimuthal wavenumber required to carry the tube to the 
boundaries of the overshoot region, $m_{\rm crit}$, decreases with increasing 
depth. 
The relationship between $\xi_r$, $m$, and the radial location of the tube is 
determined by 1) magnetic curvature force, which increases with $m$, 
and 2) the convective velocity, which decreases with depth in the overshoot 
region. 

%%%%%%%%%%%%%%%%%%%%%%%%%%%%%%%%%%%%%%%%%%%%%%%%%%%%%%%%%%%%%%%%%%%%%%%%
\subsubsection{Parameter study}
\label{sssec:par}
%%%%%%%%%%%%%%%%%%%%%%%%%%%%%%%%%%%%%%%%%%%%%%%%%%%%%%%%%%%%%%%%%%%%%%%%
\begin{figure}
	\hskip2.5mm
	\includegraphics[height=9cm,angle=90]{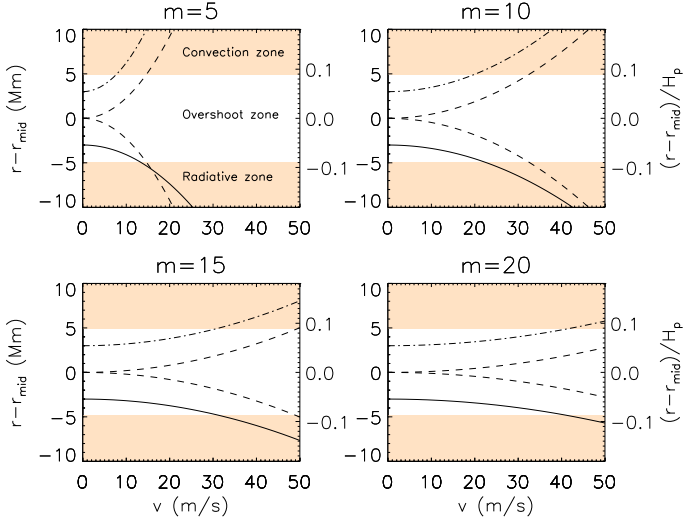} \vskip5mm
	\caption{Variation in the maximal displacement of flux tubes 
	by azimuthally periodic perpendicular flows 
	with four different azimuthal wavenumbers, $m$, as a function of 
	flow speed ($B_0=6\times 10^4$ G, $R_t=1000$ km). 
	The vertical axes denote the radial distance with respect 
	to the middle of the overshoot region 
	(on the right axes in units of the local pressure scale 
	height at the mid-overshoot region). 
	Solid lines show the maximal downward displacement of the 
        troughs of a tube starting in the bottom of the overshoot region. 
        Dashed lines show the maximal displacement for the 
        troughs and crests of tubes starting in the middle of the 
	overshoot region, and the dash-dotted lines show 
	the displacement of the crests of tubes starting near the top of 
        the overshoot region. For clarity, the effect of upward (downward) 
        flows on flux tubes near the bottom (top) of the overshoot 
	region is not shown. 
        The shaded regions denote the radiative zone below and the convection 
	zone above.}
	\label{fig:stabm_xip}
\end{figure}
The parameters describing the initial condition of the flux tube 
are the field 
strength, cross-sectional radius, initial radial position, 
and latitude. 
The parameters pertaining to the external flow are the azimuthal 
wavenumber and the velocity amplitude. 
In the following, we shall evaluate 
Eq.~(\ref{eq:sool}) and Eqs.~(\ref{eq:sool1})-(\ref{eq:sool3}) 
for given sets of parameters, in order to find the perturbation 
amplitude as functions of parameters related to the flow and the 
flux tube. 

\paragraph{Dependence on the flow velocity and the azimuthal wavenumber.}
Figure~\ref{fig:stabm_xip} shows the variation in $r-r_{\rm mid}$, 
which is the radial distance of the crest of the perturbed tube from 
the middle of the overshoot region, as a function 
of the velocity amplitude of the radial flow 
for various azimuthal wavenumbers. 
The minimum flow velocity required to advect the flux tube up to the 
convection zone or down to the radiative zone boundaries corresponds to 
the intersections of the curves with the boundaries of the overshoot 
layer. 
For a fixed flow speed and increasing $m$, it becomes 
more difficult to displace the tube crests from the equilibrium 
configuration, because the magnetic curvature force increases. 
For a given $m$, faster flows lead to larger perturbations. 
In the case for which the tube extends between the boundaries of the 
overshoot layer, the variation in buoyancy along the tube can be significant, 
because 
the superadiabaticity changes by about three orders of magnitude between 
the upper and lower boundaries 
(cf. Sect.~\ref{ssec:delta}, Fig.~\ref{fig:delta}). 
\begin{figure}
\resizebox{\hsize}{!}
{	\hskip3mm
	\includegraphics[width=6.5cm,angle=90]{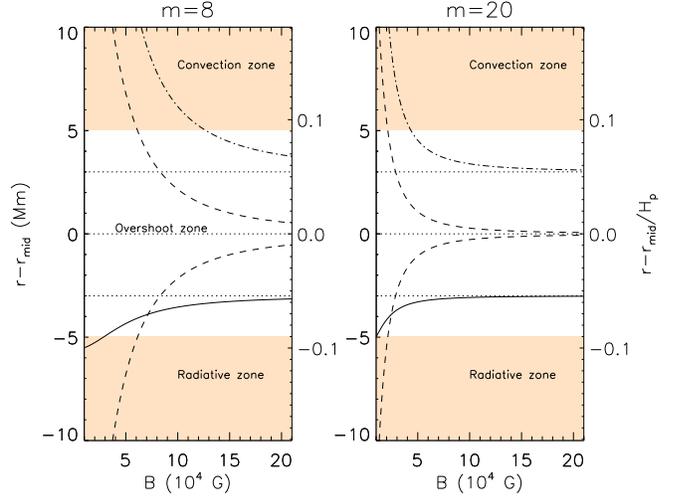}}
	\vskip5mm
	\caption{Displacement of flux tubes 
	by radial external flows that are periodic in azimuth, 
	as a function of field strength. 
	The left panel is for $m=8$, and the right panel for 
	$m=20$, to show the contrast between the effects of flows 
	with large and small wavelengths. 
	The radius, the latitude, the initial depth of the flux tubes, 
	the vertical axes, and the line styles in the plots are the same 
	as in Fig~\ref{fig:stabm_xip}. The tube is 
	subjected to an azimuthally periodic 
	flow with an amplitude of 10, 26, 31~m~s$^{-1}$ for the 
	bottom, middle, and top of the overshoot region, respectively. 
	The long dashed lines indicate the boundaries of the radiation 
	and convection zones. 
	}
	\label{fig:stabmb}
\end{figure}
At the top of the overshoot region, a small-scale deformation (a high $m$) can 
drive the flux tube into the 
convection zone proper, whereas for the bottom of the overshoot region, 
coherent downflows on a rather large scale ($m \sim 5$) are required to push the 
tube down to the radiative zone. However, the true depth of penetration 
close to the radiative zone cannot be estimated in this way, 
because the superadiabaticity 
decreases strongly, and thus the linear approach becomes inapplicable. 
Indeed, the superadiabaticity becomes strongly negative so that the stable 
stratification largely inhibits any further downward penetration 
(see Sect.~\ref{ssec:slingshot}). 

\paragraph{Dependence on the magnetic field strength.}
Figure~\ref{fig:stabmb} shows the variation in the displacement amplitude 
as a function of its field strength, for $R_{\rm t}=10^3$~km, and the 
amplitude of the azimuthally periodic perpendicular flow is 
assumed to be $14$~m~s$^{-1}$ at the bottom, $26$~m~s$^{-1}$ at the middle, 
and $31$~m~s$^{-1}$ at the top of the overshoot region, as in 
Fig.~\ref{fig:stabg}. As the azimuthal 
wavenumber of the flow is increased, the tension force resists the deformation 
of the tube more strongly, so that the perturbation weakens. 
For $B_0\gtrsim 10^5$~G and $m\gtrsim 10$, 
the deformation of 
the tube by the flow becomes smaller than $200$~km. 
For $m=20$, a perpendicular flow with a speed of $26$~m~s$^{-1}$, applied 
to a tube with $B_0=10^5$~G, located at the middle of the overshoot region, 
displaces it by an extent of about $5\times 10^{-3}H_p$ ($\sim$300~km). 
The corresponding 
relative change in superadiabaticity between the depths of the crests 
and the troughs is $\Delta\delta\lesssim 10^{-7}$ 
(see Fig.~\ref{fig:subad_pert}), so the $\delta$-effect is ineffective 
in this case. 

\begin{figure}
\resizebox{\hsize}{!}
{	\hskip3mm
	\includegraphics[width=6.5cm,angle=90]{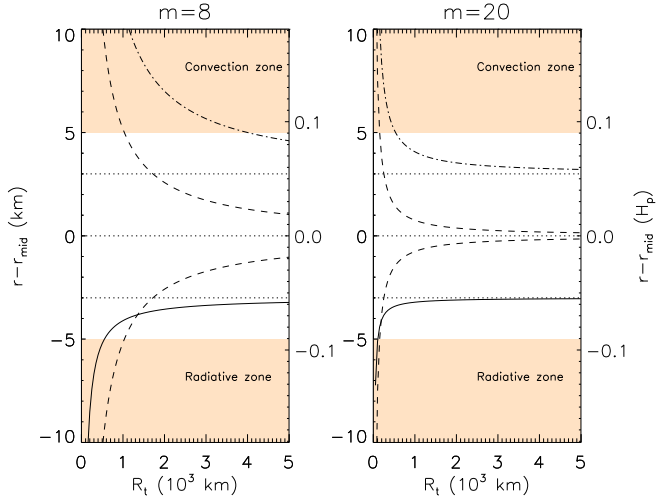}}
	\vskip5mm
	\caption{Same as Fig.~\ref{fig:stabmb}, for $B_0=6\times 10^4$~G 
	and varying tube radius. }
	\label{fig:stabmr}
\end{figure}

\paragraph{Dependence on the tube radius.}
We now set the field strength to 
$B_0=6\times10^4$~G and vary the tube 
radius, $R_{\rm t}$. The resulting functional dependence is shown in 
Fig.~\ref{fig:stabmr} for $m=8$ and $m=20$, for tubes starting 
at the three sets of depths and convective speeds chosen above for 
Fig.~\ref{fig:stabmb}. 
Because the drag force 
is inversely proportional to $R_t$, thicker tubes are less 
affected by the flow. 
For $m=20$, tubes thicker than about 2000~km in diameter do not reach the 
boundaries of the overshoot region. At first glance, this indicates that 
thicker tubes may be stored in the overshoot region for longer times than 
thinner tubes. However, 
diameters larger than about a few thousand kilometres are not relevant to 
the present context, because 
the thin flux tube approximation is not valid if the tube diameter is 
comparable to the thickness of the overshoot layer: the radial variation 
in superadiabaticity across the tube becomes non-negligible, 
inducing a strong variation in the buoyancy over the tube cross-section 
(see~Fig.~\ref{fig:delta}). 

The upper horizontal 
axis in Fig.~\ref{fig:gr-amp1} shows the perpendicular velocity amplitude 
of the external flow, leading to a given maximum radial 
displacement of the flux tube, $\Delta r=2|\xi_r|$, for $m=5$. 
Using the correspondence between $\varv_\perp$ and $|\xi_r|$ 
(Sect.~\ref{ssec:sol}), 
we can make the following estimation, using Fig.~\ref{fig:gr-amp1}. 
A toroidal flux tube with 
$\lambda_0=30^\circ$, $B_0=7\times 10^4$~G, and $R_{\rm t}=10^3$~km can be 
stored in the middle of the overshoot region for about 3 years, 
provided that the external flow velocity does not exceed about 
$10~{\rm m~s}^{-1}$. 
This semi-analytical estimate concerning the storage of a magnetic flux tube 
in the overshoot layer is tested using nonlinear 
numerical simulations involving a localised downflow, 
in Sect.~\ref{sssec:nonlinear}.

\subsection{Localised radial flow}
\label{ssec:localised}

We next consider the effects of both a localised radial flow, 
which has a finite extension in azimuth and latitude, and a 
longitudinal flow. The 
former describes, e.g., a convective downdraft penetrating into the 
overshoot layer, and the latter is required to define the initial mechanical 
equilibrium state. 

\subsubsection{Linear analysis}
\label{sssec:linear}

Before treating the nonlinear evolution of flux tubes under the combined 
effects of a localised downflow and the longitudinal flow, we develop an 
analytical approach to help us to understand the basic physics. We consider a 
toroidal flux tube subject to a localised downflow 
in the middle of the overshoot region, and describe the downward flow 
speed by a Gaussian function of the azimuth and obtain the stationary 
solution in a way similar to the one in Sect.~\ref{ssec:perlin}. 
The description of the flow field and the stationary solution 
for the displacement are given in Appendix~\ref{ssec:locflow}. 
Figure~\ref{fig:SF1T} shows the radius-azimuth 
diagrams in which the dashed curves represent the stationary 
solutions for a flow with $\varv_{\rm max}=-10$~m~s$^{-1}$ (minus sign 
means that the flow is directed downward) and various field 
strengths. For simplicity, the flux tube is located in the equatorial 
plane. 
For $B_0=10^4$~G, the downflow leads to a valley-shaped deformation 
in the flux tube. In this case, the shape of the flux tube is mostly 
determined by the azimuthal dependence of the perpendicular flow speed. 
For $B_0=3\times 10^4$~G and $B_0=4\times 10^4$~G, magnetic tension 
increasingly affects the stationary equilibrium shape of the flux tube. 
For a sufficiently strong magnetic field, a localised downflow leads 
to a deformation with a larger azimuthal extension than that of the 
downflow itself, owing to magnetic tension. 

\subsubsection{Nonlinear simulations}
\label{sssec:nonlinear}
We assume a Gaussian profile for the external flow field, 
\begin{eqnarray}
\varv_\perp = a(t)\cdot \varv_{\max} 
	\exp\left[\frac{-(r-r_{\rm m})^2}{2\sigma_r^2}
		-\frac{(\phi-\phi_{\rm m})^2}{2\sigma_\phi^2}
		-\frac{(\theta-\theta_{\rm m})^2}{2\sigma_\theta^2}\right],
\label{eq:extflow}
\end{eqnarray}
where $a(t)$ describes the time variation. 
{\rm We set \sloppy{$r_{\rm m}=5.12\times 10^{10}$~cm} } and 
$\sigma_r=5000$~km, $\theta_{\rm m}=90^\circ$, $\phi_{\rm m}=180^\circ$, 
so that the flow speed 
has its maximum value, $\varv_{\rm max}$, at the upper boundary of 
the overshoot region, 
and falls to about one fifth of its maximum in the middle of the 
overshoot region. 
The lateral extensions $\sigma_\phi$ and $\sigma_\theta$ are free 
parameters. For all the simulations, 
the flux tube is located in the middle of the overshoot layer, and 
its cross-sectional radius is $10^3$~km.

\subsubsection*{Stationary flow (SF)} 
To test whether the stationary equilibria 
predicted in Sect.~\ref{sssec:linear} 
can occur in the nonlinear case, we carried out 
numerical simulations 
by applying a localised downflow, which becomes stationary after 
a given time, $t_s=100$~days.
We consider a toroidal flux tube, which is initially in mechanical 
equilibrium. We assume that an external downflow gradually develops 
around the point $(r_{\rm m},\theta_{\rm m},\phi_{\rm m})$, 
such that the time profile in Eq.~(\ref{eq:extflow}) is assumed to be 
\begin{eqnarray}
a(t) = \left\{ \begin{array}{r@{\quad:\quad}l} t/t_s & 
{\rm if}~t\leqslant t_s \\
                              1 & {\rm if}~t > t_s.
                              \end{array} \right.
\label{eq:timeprofile}
\end{eqnarray}
\begin{figure}
\includegraphics[width=\linewidth]{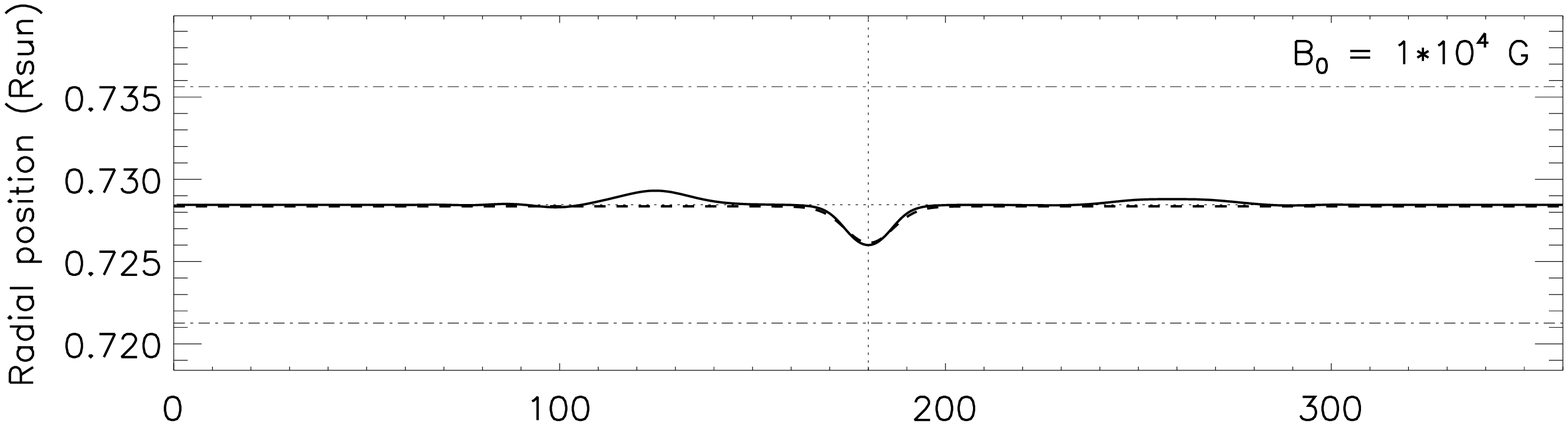} \\ \vskip-1cm
\includegraphics[width=\linewidth]{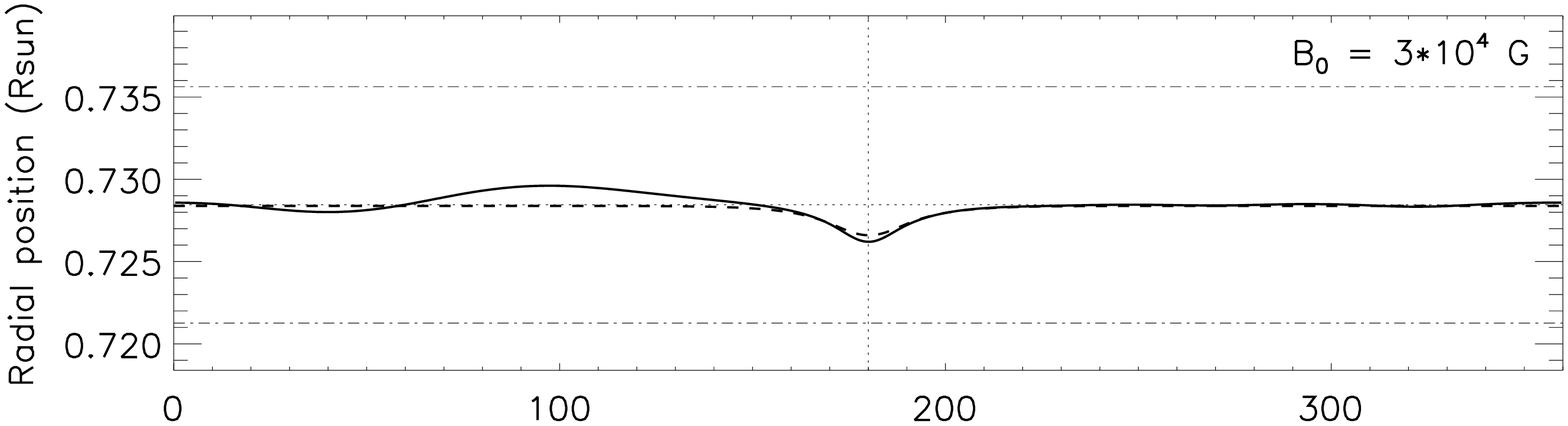} \\ \vskip-1cm
\includegraphics[width=\linewidth]{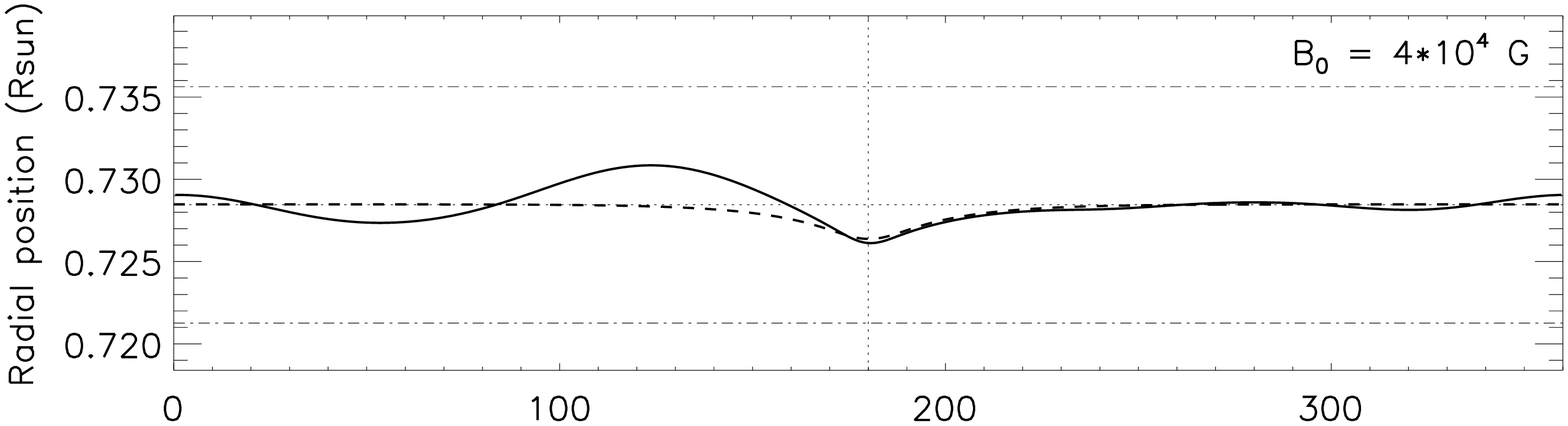} \\ \vskip-1cm
\includegraphics[width=\linewidth]{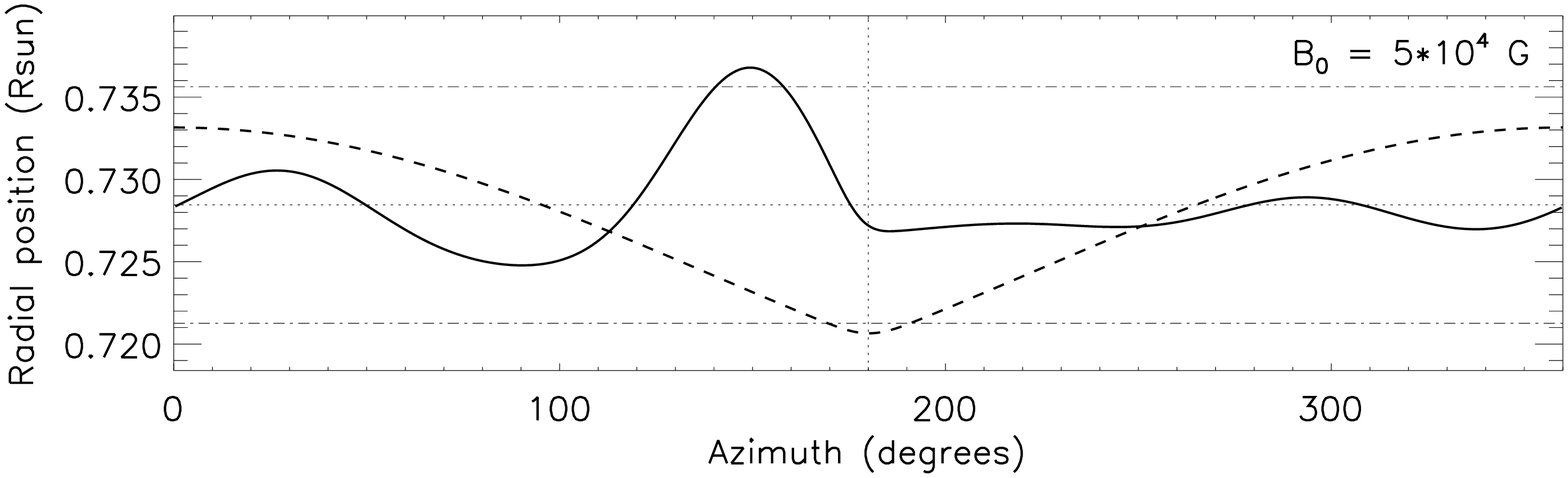}
\caption{The radial position of an equatorial flux tube as a function 
of the azimuth, as determined by linear analysis (\emph{dashed curves}, 
$\varv_{\rm max}=-10$~m~s~$^{-1}$), 
and by nonlinear simulation of the SF50 case, after 8 months 
(\emph{solid curves}). The internal flow (in the rest frame of the 
external medium) is from left to right. 
The initial field strength is different in each panel. 
The initial radial position is 
$r_0=5.07\times 10^{10}$~cm (the middle of the overshoot region). 
The dash-dotted horizontal 
lines indicate the boundaries of the overshoot region. }
\label{fig:SF1T}
\end{figure}
We present two subsets of simulations in this section: for the subset SF50, 
$\varv_{\rm max}=50$~m~s$^{-1}$, and for SF10 $\varv_{\rm max}=$10~m~s$^{-1}$. 
For each subset, the free parameters are the latitude, magnetic 
field strength, and the azimuthal and latitudinal extensions 
$(\sigma_\phi,\sigma_\theta)$ of the downflow, with the condition 
$\sigma_\phi=\sigma_\theta$. We define the ``rise time'' of an 
unstable flux tube as the time between $t=0$ and the end of the simulation 
when the tip of the rising loop arrives at about $r=0.98R_{\sun}$, where 
the thin flux tube approximation breaks down. 

Snapshots from the set SF50 at $t=8$~months for various initial 
field strengths are shown with solid curves 
in Fig.~\ref{fig:SF1T}, in comparison with the linear results 
for stationary equilibrium (Sect.~\ref{sssec:linear}). 
The direction of the internal flow along the flux tube is from left 
to right. 
The stationary solution is nearly identical to the 
nonlinear simulation snapshot for $B_0=10^4$~G. 
The broadening of the submerged part 
for higher $B_0$ (see Sect.~\ref{sssec:linear}) 
is visible in the numerical solutions. However, the simulations deviate 
from the linear results substantially on the left side of the downflow 
region, to an 
increasing extent for higher $B_0$. The azimuthally symmetric 
external downflow deforms the flux tube in an asymmetric form, 
owing to the internal flow. 
The central part of the tube reaches a dynamical equilibrium 
after about 100 days, whereas the tube as a whole never comes 
to equilibrium in any case: 
for low field strengths ($1-3\times 10^4$~G), 
the upward portion creates a transverse 
wave, which propagates leftward (retrograde). 
In the course of its subsequent evolution, the wave energy is transferred to 
lower azimuthal modes, while the submerged portion of the tube persists as 
long as the downflow speed is kept constant. 
For higher values of $B_0$, the magnetic tension force 
limits the extent of the downward displacement, whereas the excess magnetic 
buoyancy in the left wing is enhanced. This leads to a Parker-unstable 
loop for $B_0\gtrsim 5\times 10^4$~G, because the top of the loop reaches 
layers of sufficiently high superadiabaticity, and the 
density deficit in the loop top increases with time. 

\begin{figure*}
\includegraphics[width=\linewidth]{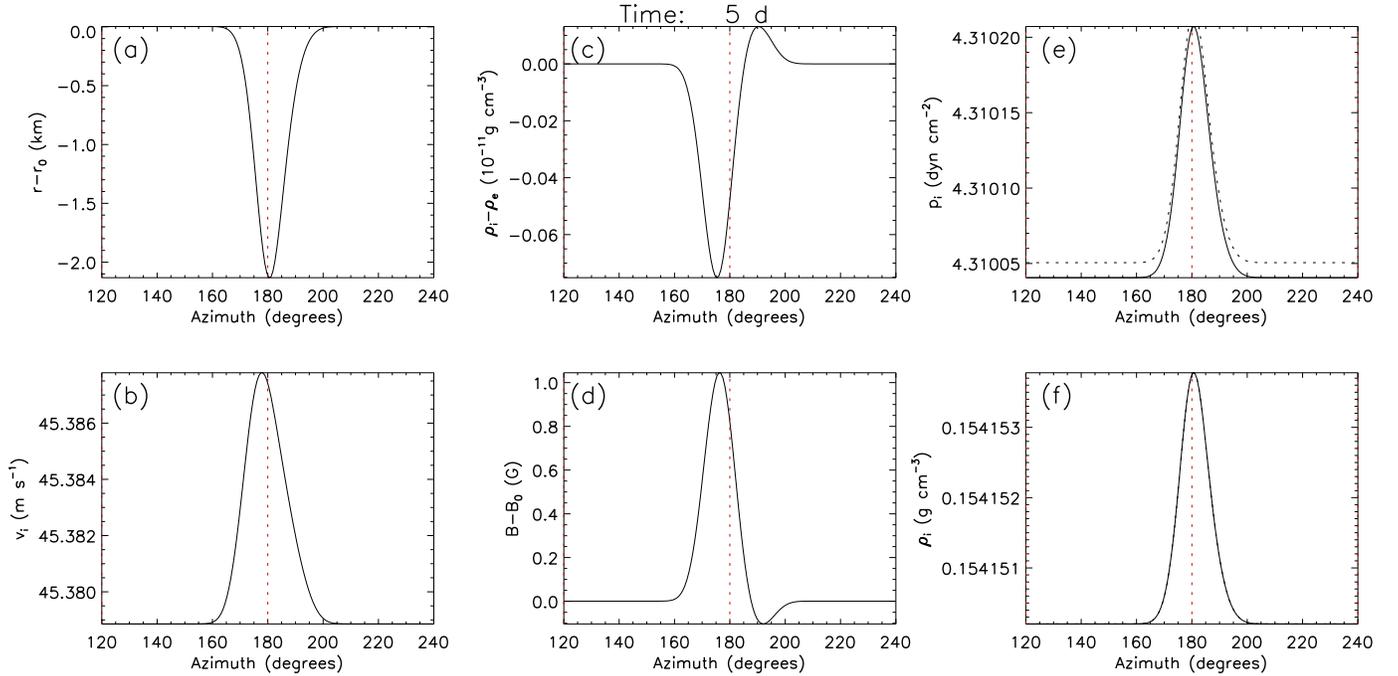} 
\caption{Physical quantities illustrating the effect of the external 
downflow, as a function of azimuth. The snapshots are taken at $t=5$~days 
of the SF50 simulation run for $B_0=5\times 10^4$~G. 
The external downflow is centred around $\phi=180^\circ$, which is 
indicated by a vertical line. We plot 
(a) the radial displacement of the tube from its initial position, 
(b) the internal flow speed,
(c) the difference between the internal and external densities,
(d) the deviation of field strength from $B_0$,
(e) the internal pressure (dashed line shows the external pressure), 
and (f) the internal density. 
}
\label{fig:phy}
\end{figure*}

To explain the physical reasons for the asymmetric 
deformation, we show in Fig.~\ref{fig:phy} 
the azimuthal profiles of various physical variables at an early 
stage of the simulation ($t=5$~days), for the SF50 case, where 
$B_0=5\times 10^4$~G. 
Once the external downflow begins to advect a part of the tube downward 
to a smaller radius (Fig.~\ref{fig:phy}a), angular momentum conservation 
leads to a local increase in the internal flow speed (b). 
The maximum internal flow speed is reached at the left wing of the 
submerged portion of the tube, because of the additional acceleration 
by the component of gravity along the tube. 
Through the right side of the submerged part, the internal flow 
decelerates back to its initial equilibrium speed. 
This asymmetric flow leads to a density deficit at the 
left part and a density excess at the right side (c). 
The resulting positive buoyancy 
in the left part increases with time and forms an upward moving 
loop, which is driven further 
by the accelerating internal flow on the right wing of the loop and 
the higher superadiabaticity at the top location (see Fig.~\ref{fig:SF1T}, 
the bottom panel). Starting from the initial phases, 
the accelerating flow also leads to a slightly lower internal pressure 
at the left part (e). 
To balance the external pressure, the magnetic field strength 
increases on the left part (d), and this 
contributes to the subsequent growth of the magnetic buoyancy instability. 
For $B_0\lesssim 5\times 10^5$~G, the instability 
does not set in, 
because the internal flow required for the initial mechanical 
equilibrium is too slow for a sufficient density deficit to develop 
on the left wing of the submerged part. 
For $B\lesssim4\times 10^4$~G, the transversal wave propagates 
leftward and quits the downflow region, well before a sufficient 
density deficit on the left wing develops. 

Table~\ref{tab:SF} presents the rise times (in cases of instability) 
of flux tubes at different latitudes, field strengths, and 
lateral extensions of the downflow, for SF50 and SF10 cases. 
Linearly Parker-unstable cases are shown in italics. 
\begin{table}%[ht]
\caption{Rise times for stationary flow simulations. }
\centering
\begin{tabular}{c c c c c}
\\
& & $\varv_{\rm max}$ = -50~m~s$^{-1}$ & & \\
\hline\hline

& & & $\tau_{\rm rise}$ (days) & \\
$\lambda_0$ (deg) & $B_0$ ($10^4$~G) &$\sigma_{\phi,\theta}=2^\circ$ & 
$\sigma_{\phi,\theta}=5^\circ$ & $\sigma_{\phi,\theta}=10^\circ$ \\
\hline
0 & 1 &	stable &	stable &	stable \\
0 & 3 &	stable &	stable &	stable \\
0 & 5 &	723 &	492 &	397 \\
0 & 6 &	920 &	324 & 	259 \\
0 & 7 &	445 &	269 &	207 \\
{\it 0} & {\it 10} &	662 &	700 &	191 \\
\hline
10 & 5 &	661 & 464 & 369 \\
10 & 7 &	439 & 265 & 204 \\
\hline
30 & 7 &	1791 &	268  & 192 \\
\hline
40 & 4 &	 stable & 2516 & 1364 \\
\hline
\\
\\
& & $\varv_{\rm max}$ = -10~m~s$^{-1}$ & & \\
\hline
\hline
& & & $\tau_{\rm rise}$ (days)& \\
$\lambda_0$ (deg) & $B_0$ (T) &$\sigma_{r,\theta}=2^\circ$ & 
$\sigma_{r,\theta}=5^\circ$ & $\sigma_{r,\theta}=10^\circ$ \\
\hline
0 & 5 &	stable & stable & stable \\
0 & 7 &	stable & stable & stable \\
{\it 0} & {\it 10} &	659 (m=1) & 608	& 567 \\
{\it 0} & {\it 12} &	147 (m=2) & 140	& 136 \\
\hline
10 & 7 & stable & stable & stable \\
\hline
30 & 7 & stable & stable & 4167 \\
\hline
\end{tabular}
\label{tab:SF}
\end{table}
For the set SF50, the cases with field strengths up to 
$5\times~10^4$~G are stable. For the linearly Parker-stable 
cases ($B<9\times 10^4$~G), greater lateral extensions of the 
downflow would destabilise the tube, because for larger $\sigma_\phi$, 
the magnetic curvature force is smaller, thus the perturbed part of the 
tube can descend deeper, increasing further the internal 
flow speed in the left wing. 
In this regime, a stronger magnetic field also leads 
to faster growth of the loop, because the internal equilibrium 
flow becomes faster, which develops the density deficit on the left wing 
more rapidly. For the linearly Parker-unstable cases 
(denoted in italics in Table~\ref{tab:SF}), the 
rise times are generally longer, because in most cases the initial 
central deformation of the tube excites standing waves with $m=5$. 
This limits the growth of the 
loop on the left side of the submerged part, and the flux tube enters 
the convection zone only after the growth of an oscillatory instability 
with $m=1$. For SF10 simulations, the downflow is too weak to trigger 
nonlinear buoyancy instability, so the linearly Parker-stable cases 
are stable in the nonlinear case, whereas for $B_0\gtrsim 10^5$~G Parker 
instability sets in, 
in accordance with the prediction of linear perturbation theory. 

\subsubsection*{Transient flow (TF)}
Downflows penetrating into the solar 
overshoot layer are probably transient, i.e., they are decelerated by 
the increasingly stable stratification on their way towards the 
radiative zone. 
To test the consequences of such a flow, we assume a Gaussian 
profile for the time variation in the downflow speed in 
Eq.~(\ref{eq:extflow}), such that 
\begin{eqnarray}
a(t) &=& \exp\left[\frac{-(t-t_{\rm m})^2}{2\sigma_t^2}\right],
\label{eq:time}
\end{eqnarray}
where $t_{\rm m}$ is the time when the maximum flow speed is reached.
The flow sets in
at $t=0$ with $\varv=10^{-3}\varv_{\rm max}$. 
The initial effect of the downflow on the flux tube is similar to 
that in the SF case. 
However, once the downflow ceases, it leaves behind transversal tube 
waves propagating along the tube, which interfere with each other. 
For unstable cases, the relative longitudinal flow leads to instability 
by frictional coupling, with the growth 
rate depending on the internal flow velocity (thus the field strength)
and the perturbation amplitude (cf. Sect.~\ref{sec:friction}). 
The difference between SF and TF is that in the former case the 
stationary flow had continued for 
a sufficiently long time to allow the formation of a buoyant loop on one 
side of the downflow region (Fig.~\ref{fig:phy}). 

We carried out a systematic survey of numerical simulations 
by varying the flow duration, $2t_{\rm m}$, and velocity amplitude, 
$\varv_{\rm max}$ in the ranges $0<|\varv_{\rm max}|<50$~m~s$^{-1}$ 
and $0<2t_{\rm m}<200$~d, for $\lambda_0=10^\circ$. 
The constants $t_{\rm m}$ and $\sigma_t$ were chosen 
such that the initial speed is of the order of $10^{-3}\varv_{\rm max}$. 
For $B_0\lesssim 5\times 10^4$~G, 
the growth time of the friction-induced 
instability is longer than the upper time limit, which 
we set to be 7 years. Instabilities with growth times longer than a 
few years are not relevant in the present context, because within that 
time the toroidal magnetic field must be amplified in the solar tachocline 
to its full strength. For $B_0=7\times 10^4$~G, 
which is about 20~kG lower 
than the Parker instability threshold for linear perturbations, the flux 
tube develops an unstable loop within about 4 years for some values of 
$(|\varv_{\rm max}|,2t_{\rm m})$. 

The evolution of the tube for $|\varv_{\rm max}|=20$~m~s$^{-1}$ and 
downflow durations of 60~days (TF60) and 180~days (TF180) is shown in 
supplementary animations (see the online appendix, Figs.~\ref{fig:TF60} 
and \ref{fig:TF180}). 
The tube rises over about 10 years for TF60, and about 3.3 years 
for TF180. In the initial stages of TF180, the evolution of the 
tube resembles the strong-field behaviour for the SF50 case 
(cf. Fig.~\ref{fig:SF1T}, bottom panel). 
 
\begin{figure}
\includegraphics[width=\linewidth]{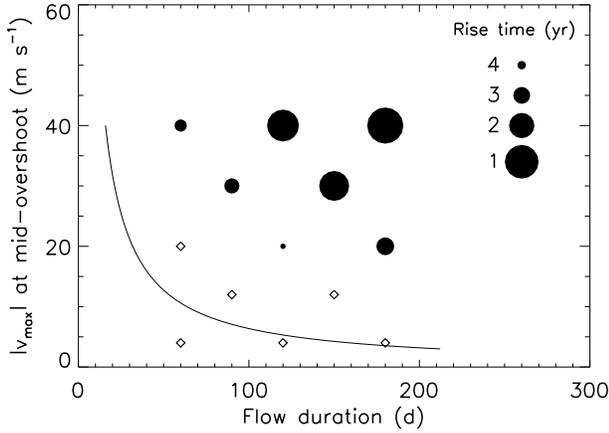}
\caption{The variation in the rise time for a linearly Parker-stable 
flux tube in the mid-overshoot region, as a function of the duration 
and maximum speed of the localised external downflow with 
$\sigma_{\phi,\theta}=5^\circ$, $\theta_{\rm m}=180^\circ$, $r_{\rm m}=512$~Mm 
(convection zone -- overshoot layer boundary) for the TF simulations. 
The field strength and latitude of the initial tube are 
$B_0=7\times 10^4$~G and $\lambda_0=10^\circ$ (cf.~Fig.~\ref{fig:stabmid}). 
The diamonds represent the ``stable'' cases, for which no emergence 
takes place before 7~years, and the size of the filled 
circles is inversely scaled with the rise time. The solid curve shows 
the convective turnover time as an indicator for the flow duration, 
estimated as $H_p/\varv_{\rm max}$.}
\label{fig:flowstab-07-10}
\end{figure}

Figure~\ref{fig:flowstab-07-10} shows the rise times as a function 
of $|\varv_{\rm max}|$ and $2t_{\rm m}$. 
Only the final month of the rise time is spent in the convection zone proper, 
the rest being spent in the overshoot region. 
A relatively fast 
downflow ($|\varv_{\rm max}|\lesssim$~40~m~s$^{-1}$) lasting less than 
about 2 months, 
or a slow downflow ($|\varv_{\rm max}|\lesssim$~20~m~s$^{-1}$) lasting 
less than about 
6 months allow the storage of the flux tube in the overshoot 
region for more than about 3~years. As a rough estimate of the flow 
duration in the overshoot region, 
we also calculated the convective turnover time, 
$H_p/\varv_{\rm max}$, in the middle of the overshoot layer. 
To lead to the formation of an unstable flux loop out of a tube with 
$B_0=7\times 10^4$~G, 
either the flow should last much longer than the turnover time 
or the flow should reach much higher speeds than the range assumed here. 

\section{Discussion}

We have investigated flow-induced instabilities of toroidal magnetic 
flux tubes in the solar overshoot region, to extend the results 
of paper~III to the nonlinear regime and to quantitatively 
estimate the effects of convective flows on the 
storage of magnetic flux in the solar overshoot region. 
Our simulations confirm the main results of the linear stability 
analysis of 
paper~III (Sect.~\ref{ssec:nonlinsims}). 
The perpendicular velocity component can be approximated generally 
in the form 
$\varv_\perp\sim|\xi_r|~~\Re{\rm e}(\omega_{\rm f})$, 
where $|\xi_r|$ is the amplitude of the radial perturbation and 
$\omega_{\rm f}$ is the eigenfrequency of the fastest growing 
unstable wave mode. Substituting this expression into Eq.~(\ref{eq:cst}), 
we obtain 
\begin{eqnarray}
\alpha \sim \frac{C_{\rm D}\varv_\perp}{\pi R_{\rm t}}
~~\propto~~
|\xi_r|~~\Re{\rm e}(\omega_{\rm f}),
\label{eq:cstext}
\end{eqnarray}
For a given value of $\Re{\rm e}(\omega_{\rm f})$, 
which is specified for a given set of 
$\delta$ (depth), $\lambda_0$, and $B_0$, $\alpha$ increases with 
increasing perturbation amplitude, as shown by numerical simulations 
in Sect.~\ref{ssec:nonlinsims} (Fig.~\ref{fig:lamslice}). 
Considering the relations obtained in Sect.~\ref{ssec:perlin}, 
a perpendicular flow with $\varv_\perp\simeq 1~{\rm m~s}^{-1}$ 
and $m=5$ in the middle of the overshoot region can 
displace a flux tube with $B=7\times10^4$~G to 
$|\xi_r| \simeq 2500$~km (cross symbols in Fig.~\ref{fig:lamslice}). 
Substituting $\varv_\perp$ into Eq.~(\ref{eq:cstext}), 
this amounts to $\alpha \simeq 3\times 10^{-7}$, which is of the 
same order as the value chosen for the full curve in 
Fig.~\ref{fig:lamslice}. 
Considering the numerical results in Sect.~\ref{ssec:nonlinsims} 
(see Fig.~\ref{fig:gr-amp1}), we conclude that an external flow 
with a velocity amplitude of 1-10~m~s$^{-1}$ perpendicular to the tube 
axis leads to instability with a growth 
time of the order of 1000 days ($\sim$2.74 years) or longer. This range 
of velocities is consistent with the estimates of \citet{vanballe82} 
for the convective velocities in the overshoot region, supporting the 
possibility of storing magnetic flux tubes that contain fluxes of 
the order of $10^{21}$~Mx. 
In a more recent study,
\citet{brummell02} carried out 3D numerical simulations of
penetrative convection, which infer penetration depths
between $0.02H_p$ and $0.11H_p$, when they extrapolate
the dimensionless numbers in their simulations to
solar values. These values are also comparable to the
radial perturbation amplitudes that we have assumed here.

The numerical experiments presented in Sect.~\ref{ssec:delta} have shown 
us that the nonlinear instability occurring in the linearly Parker-stable 
regime 
is induced mainly by the frictional coupling of the flux tube with 
its surroundings. We have also found unstable flux tube configurations, 
for which the $\delta$-effect plays a significant role in the dynamics. 
These numerical experiments were made without considering the drag force 
and in the bottom of the overshoot layer, where the radial gradient of 
superadiabaticity is steeper than in the remainder of the overshoot region. 
This result may be relevant during the final phases of the decay 
of the large-scale toroidal field in the Sun: if we assume that the toroidal 
flux at the upper layers of the overshoot region has already been removed 
to a large extent at this phase, a flux tube that forms near the bottom 
of the overshoot region can be destabilised rapidly by strong 
(possibly rare) convective 
downflows, leading to a few active regions during a solar minimum. 
We conjecture that the $\delta$-effect is not the 
main source of flux loss from the overshoot layer, with a possible 
exception in the bottom of the overshoot region, provided that 
overshooting convective flows are sufficiently strong. 

In test simulations, we have found that for perturbations larger than 
about $10^{-3}H_p$~($\sim$55~km), the growth rates start to deviate 
from the predictions of linear stability analysis \citep{afmsch95}, 
because nonlinear effects govern the dynamics and determine the growth 
rate of instability, through either the friction-induced instability 
(Sect.~\ref{ssec:nonlinsims}) or 
the $\delta$-effect (Sect.~\ref{ssec:delta}), depending on the depth 
and the longitudinal flow speed. 

In seeking possibilities of a slingshot effect that leads to 
small-scale loops in thin flux tubes, we have found that this effect 
is inefficient in removing magnetic flux from the overshoot region. 
The common result of the simulations is that a loop driven by a transient 
downflow hits the radiative zone and bounces back rapidly. However, in its 
way through the overshoot region it is strongly decelerated mainly 
by friction. 

In a parameter study surveying analytical estimates of the 
displacement amplitude as a function of field strength 
(Sect.~\ref{sssec:par}), we have found that a flow pattern with 
$m=20$ and $\varv_\perp=26$~m~s$^{-1}$ acting on a tube with 
$B_0=10^5$~G leads to a perturbation of about 300~km. 
The corresponding difference in the superadiabaticity of the 
stratification is $\Delta\delta\simeq 10^{-7}$ 
(see Fig.~\ref{fig:subad_pert}). 
Therefore, we do not expect azimuthally periodic flows 
with short wavelengths ($m\gtrsim 20$) to trigger 
flux tube instabilities for $B_0\lesssim 10^5$~G, provided that 
the displacement is less than about 300~km, in other words, 
$\varv_\perp\lesssim 26$~m~s$^{-1}$. 

After understanding the fundamental effects of azimuthally periodic 
flows on flux tubes, we have considered the case of a 
localised downflow in Sect.~\ref{ssec:localised}. 
For a given flow pattern, we have calculated the evolved states of 
flux tubes with various field strengths corresponding to linearly 
Parker-stable cases, using the steady-state approximation in conjunction 
with numerical simulations. The experiments presented in 
Fig.~\ref{fig:SF1T} can also be interpreted in terms of a toroidal field 
strength increase with time, in the rising phase of solar activity. 
As the toroidal field is amplified, localised downflows will 
have stronger disruptive effects on flux tubes, owing to the increasing 
internal flow speed, which is determined by the mechanical equilibrium 
condition. For a stationary downflow, we thus suggest that the lower 
limit to the field strength for 
which flux loops start entering the convection zone is of the 
order of $5\times 10^4$~G for the middle of the overshoot region. 

Proceeding to non-stationary, transient flows, we have set up a 
survey of simulations for $B_0=7\times 10^4$~G in the middle of the 
overshoot region for Gaussian time profiles (TF case). 
Depending on the 
dynamical properties of the downflow, we have calculated the 
evolution of a flux tube with an upper time limit of 7 years. 
However, in the first 7 years of an activity cycle, the amplification 
of the large-scale toroidal field in the tachocline cannot be neglected. 
If a flux tube with a sub-critical field strength, say 
$B_0=7\times 10^4$~G, remains in the overshoot region for a few years, 
it is likely that it will be amplified and eventually form 
a Parker-unstable loop for $B_0\sim 10^5$~G, which can lead to the 
formation of active regions with the observed properties 
in the photosphere. 

We have found that the storage of a toroidal magnetic flux tube 
with $\Phi\simeq 2\times 10^{21}$~Mx for times comparable to the 
dynamo amplification time is possible. Within the 
scope of the thin flux tube approximation and non-local mixing length 
models of the solar interior, we have not found any significant 
(hydrodynamically or magnetically induced) 
stability problem that impedes the construction of Parker-unstable 
tubes with fluxes of the order of $10^{22}$~Mx. 

We have considered the problem of storing a toroidal flux 
tube field rather than the problem of building up the magnetic field in
the first place. We have assumed that the field is amplified 
in the overshoot region, e.g., by rotational shear, 
on a timescale of a few years. The stationary equilibria of subcritical 
flux tubes ($B_0\lesssim 10^5$~G) advected by radial flows
indicate deformations comparable to the size of the overshoot region 
(see Fig.~\ref{fig:stabmb}). 
On the one hand, it may be argued that these deformations can destabilise 
the tube, e.g., by $\delta$-effect, 
at times comparable to the duration of overshooting convective flows 
(see Sect.~\ref{ssec:perlin}). 
However, nonlinear simulations (Fig.~\ref{fig:SF1T}) 
show deviations from linear estimates,
as the field strength increases towards $5\times 10^4$~G: 
the tube is destabilised as the field 
strength is \emph{increased}. 
These simulations indicate that toroidal magnetic fields can be 
stored within the overshoot region in the course of their amplification 
up to the critical field strength
($10^5$~G) required to explain general properties of active regions. 
Our results can be cross-checked
when the spatio-temporal structure of penetrative convection near the 
bottom of the convection zone is observed by helioseismology, 
or when 3D hydrodynamic simulations with realistic Reynolds numbers 
and sufficiently high resolution become available. 

\section{Conclusions}
Based on analytical steady-state approximations and nonlinear 
simulations of toroidal flux tubes in the solar convective 
overshoot region, we reach the following conclusions: 
\begin{itemize}
\item The flow instability driven by the frictional coupling 
of transversal MHD waves with external flows is nonlinear in the 
sense that the growth 
rate is a function of the initial perturbation amplitude. 
This is 
consistent with the results of the linear stability analysis of 
paper~III, in which the parameter $\alpha$ 
is proportional to the speed of the perpendicular flow and thus to 
the amplitude of the subsequent perturbation. Therefore, 
the perturbation amplitude determines the $\alpha$ parameter 
in the linear analysis. 

\item Significant buoyancy variations ($\delta$-effect) 
along magnetic flux tubes can lead to a nonlinear buoyancy instability. 
The $\delta$-effect most likely occurs in regions of large radial 
gradients of superadiabaticity such as 
the bottom of the overshoot layer, where a sufficient $\delta$ variation 
along the tube ($\Delta\delta\gtrsim 10^{-6}$) is easier to attain than 
in the upper layers. 

\item For flux tubes in the solar overshoot layer, we have 
established links between the perpendicular flow 
velocity amplitude, its spatial (azimuthal) extension, and the 
resulting radial displacement and instabilities induced by 
$\delta$-effect or longitudinal flows. 

\item To store magnetic flux with flux densities 
between $10^4$-$10^5$~G for times comparable to the dynamo 
amplification time in the convective overshoot layer, the average 
flow speed and the flow duration must not exceed about 50~m~s~$^{-1}$ 
and 100~days, respectively, and that the azimuthal extension of 
the flow is lower than about $10^\circ$. 
If we assume that these conditions prevail in the solar overshoot 
region, then magnetic fluxes of up to $10^{22}$~Mx can be 
stored within thin flux tubes during the dynamo amplification phase. 
\end{itemize}

\begin{acknowledgements}
The authors are grateful to Manfred Sch\"ussler for useful
discussions and suggestions in the course of the study, and
acknowledge the referee for the suggestions, which helped in improving the 
manuscript.
\end{acknowledgements}

%\begin{appendix}
\appendix

\section{Flux tube subject to radial flows}
\label{sec:ftradial}

The equations for a flux ring 
at an arbitrary latitude $\lambda_0$, in the limit of $\beta\gg 1$, 
are given by Ferriz-Mas \& Sch\"ussler 1995 (Sect.~4.1). 
Consider that a flow along the spherical radial direction with an amplitude 
$\varv_\perp$, is applied to a flux tube with radius $R_{\rm t}$ located at 
latitude $\lambda_0$. This flow will exert a drag force 
perpendicular to the tube axis, 
$F_{\mathrm D}=\rho_{e0}\varv_{\perp}^2 / \pi R_{\mathrm t}$, 
where $\rho_{e0}$ is the unperturbed external density. 
After a certain time, the drag force and the restoring forces of 
magnetic tension, Coriolis, and buoyancy will balance each other 
and a stationary equilibrium will be reached. 
Adopting the equation of motion for linearised perturbations 
\citep{afmsch95} for the stationary force equilibrium yields the 
inhomogeneous system of equations 
\begin{eqnarray}
% radial
2f^2\frac{\partial^2\xi_R}{\partial\phi^2} 
+ C_\phi\frac{\partial\xi_\phi}{\partial\phi} + C_R\xi_R + C_z\xi_z 
&=& \frac{\tau^2}{\rho_{e0}}F_{\rm D}(\phi)\cos\lambda_0, 
\label{eq:momeqout1} \\
\noalign{\vskip 2mm}
% azimuthal
2f^2\frac{c_T^2}{\varv_A^2}\frac{\partial^2\xi_\phi}{\partial\phi^2} 
+ A_R\frac{\partial\xi_R}{\partial\phi} + A_z\frac{\partial\xi_z}{\partial\xi_\phi}
&=& 0, \label{eq:momeqout2} \\
\noalign{\vskip 2mm}
% latitudinal
2f^2\frac{\partial^2\xi_z}{\partial\phi^2} 
+ D_\phi\frac{\partial\xi_\phi}{\partial\phi} + D_R\xi_R + D_z\xi_z 
&=& \frac{\tau^2}{\rho_{e0}}F_{\rm D}(\phi)\sin\lambda_0, \label{eq:momeqout3}
\end{eqnarray}
where 
\begin{eqnarray}
A_R = 4f\left(f-\frac{\cos\lambda_0}{2\gamma}\right), A_z = -\frac{2f}{\gamma
}\sin\lambda_0,
\nonumber \\
C_\phi = -A_R,~~~ C_R = \frac{4f}{\gamma}\cos\lambda_0 + \Delta\cos^2\lambda_0, 
\nonumber \\
C_z = \left(\frac{2f}{\gamma}+\Delta\cos\lambda_0\right)\sin\lambda_0,
\\
D_\phi = -A_z,~~~ D_R = C_z,~~~ D_z = \Delta\sin^2\lambda_0, 
\nonumber \\
\Delta = \beta\delta-\frac{2}{\gamma}\left(\frac{1}{\gamma}-\frac{1}{2}\right).
\nonumber \\
\end{eqnarray}
Here we have assumed that the magnitude of the drag force is 
comparable to a first-order perturbation. 
%%%%%%%%%%%%%%%%%%%%%%%%%%%%%%%%%%%%%%%%%%%%%%%%%%%%%%%%%%%%%%%%%%%%%%%%

\subsection{Azimuthally periodic flow}
\label{ssec:perflow}

We define the flow velocity in the form $\varv_\perp=\cos(m\phi)$ for a given 
azimuthal wavenumber $m$. The resulting drag force is given by 
\begin{eqnarray}
	F_{\rm D}(\phi,m)  &=& 
	\frac{\rho_{e0}\varv_{\perp 0}^2}{\pi R_t}\exp(im\phi),
	\label{eq:perflow}
\end{eqnarray}
where $\varv_{\perp 0}$ is the maximum value of the velocity. 
The resulting displacement is of the form 
\begin{eqnarray}
\xiup=\widehat{\xi_m}\exp(im\phi). 
\label{eq:xiup}
\end{eqnarray}
Substituting Eqs.~(\ref{eq:perflow} \& \ref{eq:xiup}) into 
Eqs.~(\ref{eq:momeqout1})-(\ref{eq:momeqout3}), we obtain 
\begin{eqnarray}
	A ~~~~~~~~~~~~~~~~~~~~~~~~~~~~~~~~~~~~~~~~~~
	\nonumber \\
	\overbrace{
	\left( \begin{array}{ccc}
	C_R-2f^2m^2	&	imC_\phi	&	C_z \\
	imA_R		&	-2f^2m^2c_T^2/\varv_A^2 &	imA_z \\
	D_R		&	imD_\phi	&	D_z-2f^2m^2
		\end{array} \right)
	}
	\left( \begin{array}{c}
	\hat{\xi}_{R} \\ 
	\hat{\xi}_{\phi} \\
	\hat{\xi}_{z}
		\end{array} \right)
	& = & \nonumber \\
	\left( \begin{array}{c}
	-\frac{\tau^2}{\pi R_{\rm t}}\cos\lambda_0\varv_{\perp 0}^2 \\ 0 \\ 
	-\frac{\tau^2}{\pi R_{\rm t}}\sin\lambda_0\varv_{\perp 0}^2
		\end{array} \right).
	\label{eq:matrix}
\end{eqnarray}
This equation can be solved in a straightforward way using 
Cramer's rule. The solution for the complex amplitudes yields 
\begin{eqnarray}
\label{eq:sool1}
\hat{\xi}_{R,m} &=& 
\frac{\tau^2m^2}{{\rm det}A}\frac{\varv_{\perp 0}^2}{\pi R_{\rm t}}
\Bigg\{\cos\lambda_0\left[2f^2\frac{c_{\rm T}^2}
{\varv_{\rm A}^2}(D_z-2f^2m^2)-A_zD_\phi\right] 
\nonumber\\
& & \hskip44pt+\sin\lambda_0\left[A_zC_\phi-2f^2\frac{c_{\rm T}^2}{\varv_{\rm A}^2}C_z\right]\Bigg\},
\\
\label{eq:sool2}
\hat{\xi}_{\phi,m} &=& 
im\frac{\tau^2}{{\rm det}A}\frac{\varv_{\perp 0}^2}{\pi R_{\rm t}}
\Bigg\{\cos\lambda_0\left[A_R(D_z-2f^2m^2)-A_zD_R\right] \nonumber\\
& & 
\hskip44pt+\sin\lambda_0\left[A_z(C_R-2f^2m^2)-A_R C_z\right]\Bigg\},
\\
\label{eq:sool3}
\hat{\xi}_{z,m} &=& 
\frac{\tau^2m^2}{{\rm det}A}\frac{\varv_{\perp 0}^2}{\pi R_{\rm t}}
\Bigg\{\cos\lambda_0\left[A_R D_\phi-2f^2\frac{c_{\rm T}^2}{\varv_{\rm A}^2}D_R\right]\nonumber\\
& & \hskip35pt+\sin\lambda_0\left[2f^2\frac{c_{\rm T}^2}{\varv_{\rm A}^2}(C_R-2f^2m^2)-A_R C_\phi\right]\Bigg\}. 
\end{eqnarray}
Substituting each component (\ref{eq:sool1})-(\ref{eq:sool3}) 
into Eq.~(\ref{eq:xiup}), we obtain 
\begin{eqnarray}
\xi_R = \Re{\rm e}(\hat{\xi}_{R,m})\cos(m\phi), \nonumber \\
\xi_\phi = -\Im{\rm m}(\hat{\xi}_{\phi,m}) \sin(m\phi), \nonumber \\
\xi_z = \Re{\rm e}(\hat{\xi}_{z,m}) \cos(m\phi), \nonumber \\ 
\label{eq:soolapp}
\end{eqnarray}

\subsection{Localised flow}
\label{ssec:locflow}

In the case of a localised perpendicular flow, we define the velocity 
and the drag force in the following forms: 
\begin{eqnarray}
	\varv_\perp(\phi) &=& \exp\left(\frac{-\sin^2\phi/2}{\sigma^2}\right), \\
	F_{\rm D}(\phi)  &=& 
	\frac{\rho_{e0}\varv_{\perp 0}^2}{\pi R_t}\sum_m \hat{g}_m\exp(im\phi),
	\label{eq:flow}
\end{eqnarray}
where $g_m=\sum_m \hat{g}_m\exp(im\phi)$ is the 
Fourier transform of the function $\varv_\perp^2(\phi)$.
For the resulting displacement, $\zetaup$, we follow the method 
described in Sect.~\ref{ssec:perflow} and write the components 
in the Fourier series representation 
\begin{eqnarray}
\zeta_R = \sum_m \hat{\xi}_{R,m} \hat{g}_m\exp(im\phi), \nonumber \\
\zeta_\phi = \sum_m \hat{\xi}_{\phi,m} \hat{g}_m\exp(im\phi), \nonumber \\
\zeta_z = \sum_m \hat{\xi}_{z,m} \hat{g}_m\exp(im\phi), \nonumber \\ 
\label{eq:soolloc1}
\end{eqnarray}
where the complex constants $\hat{\xi}_{[R,\phi,z],m}$ are given by 
Eqs.~(\ref{eq:sool1})-(\ref{eq:sool3}). 
The final solution for the components of the perturbation resulting 
from a localised perpendicular flow has the form
\begin{eqnarray}
\zeta_R = \sum_m \Re{\rm e}(\hat{\xi}_{R,m}) \left[\Re{\rm e}(\hat{g}_m)\cos(m\phi)-
	\Im{\rm m}(\hat{g}_m)\sin(m\phi)\right], \nonumber \\
\zeta_\phi = -\sum_m \Im{\rm m}(\hat{\xi}_{\phi,m}) \left[\Im{\rm m}(\hat{g}_m)\cos(m\phi)+
	\Re{\rm e}(\hat{g}_m)\sin(m\phi)\right], \nonumber \\
\zeta_z = \sum_m \Re{\rm e}(\hat{\xi}_{z,m}) \left[\Re{\rm e}(\hat{g}_m)\cos(m\phi)-
	\Im{\rm m}(\hat{g}_m)\sin(m\phi)\right]. \nonumber \\
\label{eq:soolloc2}
\end{eqnarray}

\section{The stationary equilibrium approximation }
\label{sec:assumption}

\subsection{Comparison of azimuthal and radial effects }
\label{ssec:justify}

Suppose that a toroidal flux tube is in mechanical equilibrium. 
Then apply a localised external flow in the spherical radial direction. 
The parts of the tube affected by the flow are advected 
in the direction of the flow, until the drag force is balanced by the 
restoring forces of buoyancy and magnetic tension. 
To justify ignoring any displacement of mass elements 
along the tube, a necessary condition is 
that the time for the radial displacement, $\tau_r$, is short compared 
to that of the displacement along the tube axis 
(tangential perturbation, $\tau_\phi$~), so that 
\begin{eqnarray}
\tau_\phi \equiv 
\frac{\pi r_0}{m|\dot{\xi}_\phi|} \gg \left|\frac{\xi_R}{\varv_R}\right|
\equiv \tau_R, 
\label{eq:req}
\end{eqnarray}
where $r_0$ is the radial coordinate of the equilibrium tube, $\dot{\xi}_\phi$ 
is the propagation velocity of the perturbation along the tube, and 
$\varv_R$ is the maximum flow speed, and $\dot{\xi}_\phi$ can be determined 
from the solution of the eigenvector problem using the linearised equation 
of motion for first-order perturbations (see Sect.~\ref{ssec:eigenvec}). 
The left-hand side of Eq.~(\ref{eq:req}) represents a timescale 
for the re-establishment of hydrostatic equilibrium, and the right-hand side 
represents the time it takes for a dynamical balance between the drag 
force and the restoring forces is reached. 

As an example, we consider an equatorial flux tube, in the middle of 
the overshoot layer. The external flow is assumed to be localised 
(Eq.~\ref{eq:flow}) and 
downwards with a maximum speed of $10$~m~s$^{-1}$. 
This is the same flow configuration as that described in 
Sects.~\ref{sssec:linear} and \ref{ssec:locflow}. 
For simplicity, we disregard the effects 
of longitudinal wave modes and the presence of azimuthal flow, 
which is required by the mechanical equilibrium condition. 
Table~\ref{tab:timesc} shows the values for both sides of the inequality 
(\ref{eq:req}), 
as a function of the azimuthal wavenumber of the external flow, 
for $B_0=6\times 10^4$~G. We also give the values of the downward 
displacement amplitude, $\xi_R$, produced by the external downflow. 
For comparison, numerical simulations were performed by setting up 
azimuthally periodic flow configurations for various azimuthal 
wavenumbers. In each case, the flow sets in at $t=0$ and reaches its 
full strength ($|\varv_{\rm max}|=10$~m~s$^{-1}$) in 30 days. 
The tabulated values are the maximum displacement at $t=240$~days, 
for which the drag force exerted by the external flow is already 
balanced by buoyancy and magnetic tension. 
The variation in displacement amplitude as a function of $m$ 
for azimuthally periodic flows shows general agreement with that 
for the linearly estimated 
amplitudes of each azimuthal wave mode corresponding to a localised 
downflow. 

The ratio $\tau_\phi/\tau_R$ decreases for smaller $m$. When 
$\tau_\phi$ becomes comparable to $\tau_R$ (as the ratio falls 
below about 10), the azimuthal perturbation might have a significant effect 
on the density distribution along the tube, such that our assumption of a 
stationary equilibrium between the normal forces becomes inaccurate. 

\begin{table}[ht]
\centering
\caption{Characteristic times for radial and azimuthal displacements.}
\begin{tabular}{c c c c c c}
\\
\hline
$m$  &	$\xi_R$ (linear) & $\xi_R$ (num.sim.)  &  $\tau_R$ (d) & $\tau_\phi$ (d) & $\tau_\phi/\tau_R$ (d) \\
& (km) & (km) & & & \\
\hline\hline
3  &	14077  & 12180 & 16.3  &	85  &	5.2 \\
4  &	4347  & 4037 & 5.0  &	202  &  40 \\
5  &	2302  &	2366 & 2.7  &	290  &	109 \\
6  &	1461   & 1531 &	1.7  &	358  &	211 \\
7  &	1021  &	1044 & 1.2  &	408  &	345	 \\
8  &	757  & 731 &	0.9  &	445  &	509 \\
\hline
\end{tabular}
\label{tab:timesc}
\end{table}

\subsection{Flux tube in the equatorial plane: eigenvector problem}
\label{ssec:eigenvec}

For simplicity, we consider a flux tube in 
the equatorial plane, to obtain eigenvectors for the wave-like solutions 
of the perturbed flux tube. In this case, the momentum equation for linear perturbations 
is simpler than for a non-equatorial flux tube, 
and therefore it is simpler to make a physical interpretation. 

To obtain relations between the flow velocity and the resulting 
displacement, we consider the linearised equations of motion for 
perturbations 
$(\xi_R,\xi_\phi,\xi_\lambda)$, where $R$ is the distance from the rotation 
axis, $\phi$ is the longitude, and $\lambda$ is the latitude. Here, only radial 
and azimuthal components (perturbations in the $(R,\phi)$ plane) are considered, 
because the equation for the latitudinal component is decoupled in the equatorial 
plane. Following Ferriz Mas \& Sch\"ussler (1993), the linearised equations 
of motion for the perturbations in the $(R,\phi)$ plane read 
\begin{eqnarray}
	\tau^2\left(\ddot{\xi}_R-2\Omega\dot{\xi}_\phi\right) & = &
	2f^2\frac{\partial^2\xi_R}{\partial\phi_0^2} - 
	4f\left(f-\frac{1}{2\gamma}\right)\frac{\partial\xi_\phi}{\partial\phi_0} +
	T\xi_R,
	\label{eq:r0} \\
	\noalign{\vskip 2mm}
	\tau^2\left(\ddot{\xi}_\phi+2\Omega\dot{\xi}_R\right) & = &
	2f^2\frac{\partial^2\xi_\phi}{\partial\phi_0^2} + 
	4f\left(f-\frac{1}{2\gamma}\right)\frac{\partial\xi_R}{\partial\phi_0}, 
	\label{eq:fi0} 
\end{eqnarray}
where 
\begin{eqnarray}
	T & = &
	2(\sigma-1)f^2 +
	\frac{4}{\gamma}f - 
	\frac{2}{\gamma}\left(\frac{1}{\gamma}-\frac{1}{2}\right) \nonumber\\
& &	+ \beta\delta + 
	\tau^2(\sigma-1)(\Omega_\mathrm{e0}^2-\Omega^2),
	\label{eq:T} \\
	\tau	& \equiv & \left(\frac{\beta H_p}{g_0}\right)^{1/2}
		 = \sqrt{2}\frac{H_p}{\varv_A},
	\label{eq:tau} \\
	f	& \equiv & \frac{H_p}{R_0} = \frac{p_{i0}}{g_0\rho_{i0}R_0}, 
	\label{eq:f} \\
	\beta	& \equiv & \frac{8\pi p_{i0}}{B_0^2}. 
	\label{eq:beta}
\end{eqnarray}
The meanings of the various symbols in the above equations are as follows: 
$H_p$ is the local pressure scale height, $\delta$ is the superadiabaticity, 
$p_{i0}$ and $\rho_{i0}$ are 
the equilibrium values for the pressure and density inside the tube, respectively, 
$r_0$ is the 
spherical radial position of the equilibrium tube, $\Omega$ and $\Omega_{e0}$ 
are, respectively, the angular velocities of 
the plasma in the tube and the external medium, $\varv_A$ is the 
Alfv\'{e}n velocity, $B_0$ is the magnetic field strength, and $g_0$ the 
gravitational acceleration with a radial dependence 
$g \propto r^{\hskip0.75mm\sigma}$, where $\sigma$ is taken to be $-2$, 
which means that the small contribution of the mass inside the convection zone 
is neglected. 
The substitution of the \emph{ansatz}
\begin{eqnarray}
	{\bf \xi}_m &=& \widehat{\xi}_m\exp(i\omega t + im\phi).
\label{eq:ansatz0}
\end{eqnarray}
into Eqs.~(\ref{eq:r0})-(\ref{eq:fi0}) leads to the system of equations 
\begin{eqnarray}
	(2f^2m^2-T-\tilde{\omega}^2)\hat{\xi}_R 
	+ i\left[4fm\left(f-\frac{1}{2\gamma}\right) 
	- 2\tilde{\Omega}\tilde{\omega}\right]\hat{\xi}_\phi = 0 \\
	-i\left[4fm\left(f-\frac{1}{2\gamma}\right) 
	- 2\tilde{\Omega}\tilde{\omega}\right]\hat{\xi}_R 
	+ (2f^2m^2 - \tilde{\omega}^2)\hat{\xi}_\phi  = 0.
	\label{eq:eigenvecsys}
\end{eqnarray}
For the existence of non-trivial solutions, the characteristic 
determinant must vanish. This yields the dispersion relation, 
which is a 4th-order polynomial in terms of the eigenfrequency. 
The eigenvector in the azimuthal direction is determined by one of the 
equations in the system 
(\ref{eq:eigenvecsys}): 
\begin{eqnarray}
\hat{\xi_\phi} = i \left[\frac{2\tilde{\Omega}\tilde{\omega}-4fm(f-1/2\gamma)}
		{\tilde{\omega}^2-2f^2m^2}\right]\hat{\xi_R},
\label{eq:eigenvec}
\end{eqnarray}
from which we obtain, by using the \emph{ansatz} Eq.~(\ref{eq:ansatz0}), the velocity
\begin{eqnarray}
\hat{\dot{\xi}_\phi} = i\omega_m \hat{\xi_\phi}
\label{eq:eigenvel}
\end{eqnarray}
of the azimuthal perturbation, as an estimate of the rate at which the hydrostatic 
equilibrium along the perturbed flux tube is established. We use this 
result in evaluating Eq.~(\ref{eq:req}), to check for the validity 
of the assumption of stationary equilibrium 
of a flux tube being displaced by an external flow. 

%\end{appendix}

%\pagebreak
%\newpage

\bibliographystyle{aa}
\bibliography{12816pap}

\Online

%\begin{appendix}
%\appendix

%\section*{Online appendix}
\section{Animations}

We provide three animated GIF files showing the simulated evolution 
of perturbed flux tubes, available on-line. 

\begin{figure}
\includegraphics[width=\linewidth]{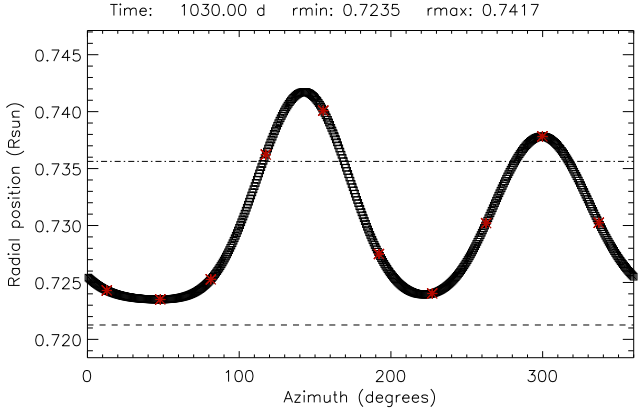}
\caption{The shape of a flux tube with 
$B_0=7\times 10^4$~G, $\lambda_0=30^\circ$ and $\Delta r=5508$~km 
at $t=1030$~d, in the $(r,\phi)$ plane. 
The horizontal lines mark the boundaries of the overshoot region.}
\label{fig:frict_inst}
\end{figure}

The animation {\tt frict\_inst.gif} shows 
the final phases of the development of the friction-induced instability 
in the overshoot region, between $t=700$~d and $t=1100$~d 
(Sect.~\ref{ssec:nonlinsims}). 
A snapshot from the animation is shown in Fig.~\ref{fig:frict_inst}. 
The asterisk signs represent selected 
mass elements, which all move rightwards in the initial phases, 
owing to the internal equilibrium flow. We assume that 
$B_0=7\times 10^4$~G, $\lambda_0=30^\circ$, and the initial radial 
perturbation amplitude is $\Delta r=5508$~km. 
The initial location is the middle of the overshoot region. 

\begin{figure}
\includegraphics[width=\linewidth]{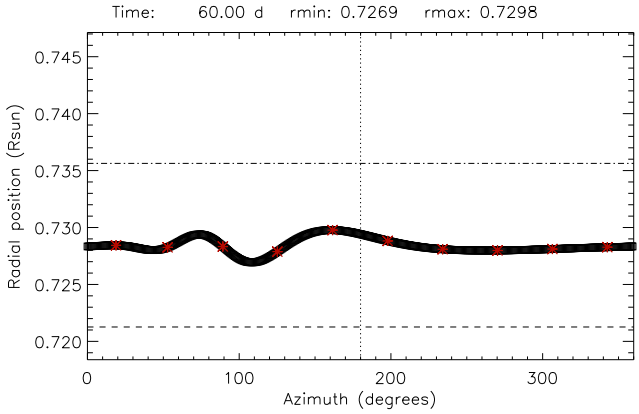}
\caption{The shape of a flux tube with $B_0=7\times 10^4$~G, 
$\lambda_0=10^\circ$, at $t=60$~d, when the transient downflow 
($\varv_{\rm max}=20$~m~s$^{-1}$) ceases. The horizontal lines mark 
the boundaries of the overshoot region 
and the vertical line denote the azimuthal location of the centre 
of the downflow.}
\label{fig:TF60}
\end{figure}

\begin{figure}
\includegraphics[width=\linewidth]{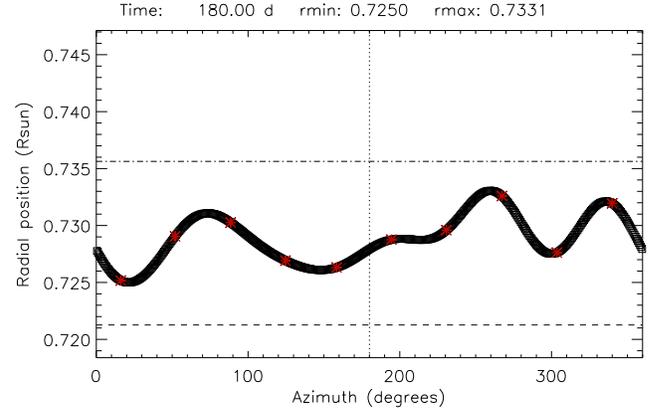}
\caption{Same as Fig.~\ref{fig:TF60}, at $t=180$~d, when the transient 
downflow ceases.}
\label{fig:TF180}
\end{figure}

The animation files {\tt TF60.gif} and {\tt TF180.gif} show the evolution 
of a flux tube subject to a radial downflow with a duration of 60 and 
120 days, respectively, until $t=300$~days (Sect.~\ref{sssec:nonlinear}). 
Two snapshots are shown in Figs.~\ref{fig:TF60} 
and \ref{fig:TF180}, corresponding to the time when the downflow ceases 
in each case, i.e., at $t=60$~d and $t=180$~d. Note that for the longer-duration 
flow, the flux tube is radially more disturbed when the action of the downflow is 
finished. 


%\end{appendix}

\end{document}